\pdfoutput=1
\documentclass[twocolumn,floatfix,tighten,times]{aastex631}

\usepackage{graphicx}    

\usepackage{float} 

\usepackage{dcolumn}     
\usepackage{bm}          

\usepackage{amsmath} 
\usepackage{amssymb} 
\usepackage{amsfonts} 
\usepackage{latexsym}
\usepackage{enumitem} 

\usepackage{booktabs} 

\usepackage{natbib}

\usepackage[normalem]{ulem}

\usepackage[english]{babel}
\usepackage[expansion=all,babel]{microtype}

\usepackage{cleveref}
\usepackage{mathtools}

\usepackage{multirow}

\usepackage{rotating}

\newcommand{\sat}[1]{{#1}_{\mathrm{sat}}}
\newcommand{\sym}[1]{{#1}_{\mathrm{sym}}}

\newcommand{\eff}[1]{{#1}_{\mathrm{eff}}}

\usepackage{txfonts}

\newcommand{\be}{\begin{equation}}
\newcommand{\ee}{\end{equation}}
\newcommand{\ba}{\begin{eqnarray}}
\newcommand{\ea}{\end{eqnarray}}
   
\newcommand{\Msun}{M_\odot}

\makeatletter
	\newcommand{\vast}{\bBigg@{2.85}}
\makeatother

\shorttitle{General-purpose EOSs for astrophysical simulations}
\shortauthors{Raduta \& Beznogov}
\begin{document}

\title{New {\em ab initio} constrained extended Skyrme equations of state for simulations of neutron stars, supernovae and binary mergers: I. Subsaturation density domain}

\author[0000-0001-8421-2040]{Adriana R. Raduta}
\affiliation{National Institute for Physics and Nuclear Engineering (IFIN-HH), RO-077125 Bucharest, Romania}
\email{araduta@nipne.ro}

\author[0000-0002-7326-7270]{Mikhail V. Beznogov}
\affiliation{National Institute for Physics and Nuclear Engineering (IFIN-HH), RO-077125 Bucharest, Romania}
\email{mikhail.beznogov@nipne.ro}

\date{\today}

\begin{abstract}
In numerical simulations of core-collapse supernova and binary neutron stars mergers, information about the energetics and composition of matter is implemented via external tables covering the huge ranges of thermodynamic conditions explored during the astrophysical evolution. More than 120 general-purpose equation of state (EOS) tables have been contributed so far. Unfortunately, not all of them comply with current constraints from theoretical and experimental nuclear physics and astrophysical observations of neutron stars. Systematic investigations of the role that dense matter properties play in the evolution of these astrophysical phenomena require that more EOS tables are provided. We build a set of general-purpose EOS tables. At zero temperature, they comply with all currently accepted constraints, including {\em ab initio} chiral effective field theory calculations of pure neutron matter. This set is designed to explore a wide variety of the behaviors of the effective masses as functions of density, which is reflected into a wide range of thermal behaviors. We employ Brussels extended Skyrme interactions generated by means of Bayesian inference techniques. An extended nuclear statistical equilibrium model is developed for modeling sub-saturated inhomogeneous nuclear matter (NM). Here, we study the properties of sub-saturated inhomogeneous NM over wide ranges of density, temperature, and proton fraction. We analyze in detail the mechanisms of transition to homogeneous matter and estimate the transition density. Our key results include a thick layer of neutron rich isotopes of He or H in the inner crusts of neo-neutron stars, significant abundance of exotic isotopes of H and He in warm and neutron-rich matter and a detailed study of the thermodynamic stability of cold stellar matter. The EOS tables are publicly available in the \textsc{CompOSE} online database.
\end{abstract}



\section{Introduction}
\label{sec:Intro}

Numerical simulations of core-collapse supernova (CCSN) \citep{Janka_PhysRep_2007,Mezzacappa2015,Schneider_PRC_2017,Connor2018ApJ,Burrows2020MNRAS} and binary neutron stars mergers (BNS) \citep{Shibata_11,Rosswog_15,Baiotti_2017,Endrizzi_PRD_2018,Ruiz2020,Prakash_PRD_2021,Most_2022}, evolution of proto-neutron stars (PNS) \citep{Pons_ApJ_1999,Pascal_MNRAS_2022} and formation of black holes (BH) in failed CCSN \citep{Sumiyoshi_2007,Fischer_2009,OConnor_2011,hempel12} require accurate microphysics input on particle composition,
thermodynamic properties and reaction rates~\citep{Oertel_RMP_2017}. Composition and thermodynamic information is customarily provided in tabular form of the so-called general-purpose equation of state (EOS) tables, over wide domains of temperature ($0 \leq T \leq 100~\mathrm{MeV}$), density ($10^{-14}~\mathrm{fm}^{-3} \leq n_B \leq 1.5~\mathrm{fm}^{-3}$) and electron fraction ($0 \leq Y_e=n_e/n_B \leq 0.6$) and with a mesh fine enough to allow interpolation and computation of additional thermodynamic state variables by differentiation~\citep{Typel_EPJA_2022}. Some EOS tables also contain information on specific microscopic quantities, e.g., effective masses and interaction potentials and neutrino opacities.

Increased interest in the physics of neutron stars (NS) and BNS mergers, largely motivated by the huge progress done by the multi-messenger astronomy over the last decade, made that a significant number of general-purpose EOS tables has been built and recently made available for numerical simulations.
The \textsc{CompOSE} online repository (\url{https://compose.obspm.fr/}) hosts more than 120 such tables and is still expanding.
This collection reflects the wealth of assumptions regarding the particle degrees of freedom, theoretical approaches, and effective interactions.
It is also illustrative of the uncertainties that affect states of matter that are impossible to produce and study in terrestrial laboratories.

Systematic studies of CCSN~\citep{Schneider_PRC_2019b,Yasin_PRL_2020,Andersen_ApJ_2021}, BNS mergers~\citep{Fields_ApJL_2023,Raithel_PRD_2023} and stellar BH formation~\citep{Schneider_ApJ_2020} demonstrated
that the evolution of these phenomena is linked to the behavior of dense matter, with
the nucleon effective mass ($\eff{m}$) playing the most important role.
During the post-bounce evolution, $\eff{m}$ impacts the properties of both PNS and neutrinosphere.
\cite{Schneider_PRC_2019b,Yasin_PRL_2020} showed that large values of $\eff{m}$ result in higher (lower) central densities (temperatures) in the PNS core and lower PNS radii. Clear positive (negative) correlations have been identified also between $\eff{m}$ and the neutrinosphere's temperature and proton fraction (density and radius) as well as neutrino energies and luminosities~\citep{Schneider_PRC_2019b}. \cite{Schneider_ApJ_2020} proved that, for a failed CCSN, the collapse into a BH occurs earlier for higher values of $\eff{m}$.

Studies of CCSN also indicate some sensitivity to the composition of sub-saturated ($n_B \leq \sat{n} \approx 0.16~\mathrm{fm}^{-3} \approx 2 \cdot 10^{14} \mathrm{g/cm^3}$) nuclear matter (NM).
An example in this sense is offered by \cite{Pascal_PRC_2020}, who showed that the Single Nucleus Approximation (SNA) and Nuclear Statistical Equilibrium (NSE) versions of the same EOS lead to different neutronization degrees of stellar matter during the in-fall stage of a CCSN and different neutrino mean free paths.
\cite{Pascal_PRC_2020} explain these discrepancies in terms of differences between the electron capture rates of the unique representative nucleus in SNA and the ensemble of nuclei in the NSE distribution, respectively.

The role of light clusters for CCSN and neutrino driven winds was considered by \cite{Sumiyoshi_PRC_2008} and \cite{Arcones_PRC_2008}, respectively. According to \cite{Sumiyoshi_PRC_2008}, abundant production of light clusters affects the effectiveness of neutrino reheating of the shock wave, as well as the structure and composition of PNS.
According to \cite{Arcones_PRC_2008}, light nuclei have a small impact on the average energy of the emitted
electron neutrinos, but are significant for the average energy of antineutrinos.


At temperatures around $0.5~\mathrm{MeV}$, properties of the sub-saturated NM also govern the processes of formation and crystallization of NS crusts, thus determining their final composition, thermal and electrical conductivities~\citep{Carreau_AA_2020,Fantina_AA_2020,Potekhin_AA_2021}.


The main aim of this paper is to propose a new set of general-purpose EOSs which, upon input in numerical astrophysical simulations, will contribute to a better understanding of the role that the nuclear EOS plays. 
All these EOSs employ Brussels extended Skyrme effective interactions~\citep{Chamel_PRC_2009}; they are built within a Bayesian inference of the EOS of dense matter~\citep{Beznogov_PRC_2024} and selected such that they feature distinct thermal behaviors in the supra-saturation regime.
The sub-saturation density domain, where NM is inhomogeneous, is treated within the NSE approximation.
Homogeneous NM is treated within the non-relativistic mean-field model~\citep{Negele_PRC_1972,Vautherin1996}. 

The high convergence performance of our new NSE code made it possible to approach the nuclear saturation density for a wide range of electron fractions and temperatures. 
In the first place, this allowed us to study in detail the mechanisms of transitioning to homogeneous matter.
We show that, depending on $T$ and $Y_p$, the transition occurs either due to the gas of nuclei or the gas of self-interacting unbound nucleons filling the entire volume. 
Special attention is also devoted to the stability of stellar matter as a mixture of nuclear clusters, unbound nucleons, and a gas of electrons. 
We demonstrate that the density-driven instabilities featured by homogeneous NM do not manifest in clusterized NM matter with electrons. 

The paper is organized as follows. In Sec.~\ref{sec:NSE}, we present the NSE formalism along with how it is implemented in our model.
The properties of a number of newly proposed Brussels extended Skyrme effective interactions are also described. 
Sec.~\ref{sec:Transition} discusses the various mechanisms by which the transition to homogeneous matter occurs in our model.
The thermodynamic stability of a cold mixture of nuclei, nucleons, and electrons is addressed in Sec.~\ref{sec:Stability}; special attention is given to the stability of beta-equilibrated matter.
In Sec.~\ref{sec:Results}, the results of our NSE model are investigated over wide domains of temperature, electron fraction, and density.
The composition of matter is analyzed in terms of three generic components, heavy and light nuclei, and unbound nucleons.
Then, the density dependence of a few thermodynamic state variables is analyzed for fixed values of $T$ ($Y_p$) and variable values of $Y_p$ ($T$).
NSE results at low temperatures are used to infer the composition of NS crusts.
The conclusions are drawn in Sec.~\ref{sec:Conclusions}.


\section{Extended Nuclear Statistical Equilibrium}
\label{sec:NSE}

For densities lower than the normal nuclear saturation density $\sat{n}$ and
temperatures lower than a dozen MeV, stellar matter consists of a mixture of nuclei, nucleons, leptons, and photons.
Charged leptons are essential for compensating for the positive charge of the protons.
Assuming that charge neutrality holds locally, $n_p=\bar n_e+ \bar n_{\mu}$,
where $\bar n_{e/\mu}=n_{e^-/\mu^-}-n_{e^+/\mu^+}$ stands for the net number of electrons and muons
and $n_p$, $n_{e^-}$, $n_{\mu^-}$, $n_{e^+}$, $n_{\mu^+}$ represent the number densities of protons, electrons, muons, positrons and anti-muons.
The only significant interaction between electrons and protons is the electrostatic one, for which the Wigner-Seitz approximation~\citep{Baym_NPA_1971} is customarily employed.
The electrons are assumed to be uniformly distributed throughout the volume, and the electron gas is treated as an ideal Fermi gas.
Photons are thermalized and are described as an ideal Bose gas.
Neutrinos are not necessarily in equilibrium with the rest of matter and are disregarded here.
For simplicity, we shall also disregard muons. 
Accounting for them is straightforward. 

For very low temperatures, the nuclear composition of stellar matter is decided upon a reaction network.
As the temperature increases, more nuclear reactions become possible and their cross sections increase.
At $T \approx 4-5~\mathrm{GK}$, all nuclear reactions proceed at the same rate as their inverses, which makes that the species' abundances are not decided any longer by the reaction rates~\citep{Wiescher2018}. 
This is the so-called NSE regime, where species abundances are determined by nuclear binding energies and internal partition functions.

In the evolution of the CCSN, PNS, and BNS mergers, densities higher than $\sat{n}$ and temperatures up to several tens of MeV are reached.
The obvious need for a unified treatment of the sub- and supra-saturation density domains urged the community to further develop the NSE approach to the point of approaching the transition to homogeneous matter and more appropriately dealing with configurations dominated by unbound nucleons.
As such, the so-called extended NSE (eNSE) approach arose.
In contrast with the standard NSE, where nuclei are treated as an ideal gas, the extended version accounts for interactions among unbound nucleons on the one hand and among unbound nucleons and nuclei on the other hand.
The ensemble of self-interacting unbound nucleons is treated within the mean-field theory of NM, which is also used to describe the matter in the supra-saturation regime.
Interactions among nuclei and with unbound nucleons are phenomenologically implemented via the excluded volume.
As we shall see in the following, this schematic treatment turns out to be sufficient for a reasonable description of matter with densities in the proximity of the transition to homogeneous matter. 

Several eNSE models have been proposed in the last 15 years. The models by \cite{Hempel_NPA_2010}, \cite{Pais_WS_2017,Typel_JPG_2018} and \cite{Furusawa_ApJ_2011,Furusawa_ApJ_2013,Furusawa_JPG_2017,Furusawa_NPA_2017} treat the gas of unbound self-interacting nucleons within the covariant density functional theory of NM while the models by \cite{Raduta_PRC_2010,Gulminelli_PRC_2015,Raduta_NPA_2019} employ the non-relativistic mean-field model of NM. 
While in all these models nuclei are assumed to form an ideal gas, their description varies significantly from one implementation to another.
For the binding energies, either experimental values, evaluations by dedicated models, or liquid drop parametrizations have been employed.
Temperature-induced excitations are accounted for via various level density expressions; the assumptions made in regard to the upper limit of the excitation energy range from the minimum between neutron and proton separation energies up to infinity.
The generalized relativistic mean-field model by \cite{Pais_WS_2017,Typel_JPG_2018} implements in-medium modifications of the nuclear energy functional via parametrized mass shifts derived from a quantum statistical model~\citep{Typel_PRC_2010}.
These shifts mimic the Pauli blocking of clusters' states by the nucleons in the medium and guarantee that nuclei dissolve in high-density media.
They are so efficient in accounting for interactions among nuclei and nucleons that they render the excluded volume approximation used by other eNSE models useless.
Temperature and in-medium effects on nuclear energy functional are phenomenologically accounted for also by \cite{Furusawa_ApJ_2011,Furusawa_ApJ_2013,Furusawa_JPG_2017,Furusawa_NPA_2017}.
Of course, many nucleon-nucleon effective interactions have been used, which is mainly reflected in different behaviors of matter with densities higher than the saturation density and temperatures higher than $10-20~\mathrm{MeV}$.  
For a comparative analysis of various NSE models, see \cite{Buyukcizmeci_NPA_2013,Raduta_EPJA_2021}.

In the following, we describe the model proposed in this work.

\subsection{Formalism}
\label{ssec:Formalism}

Here, the canonical ensemble is used. The controlled quantities are the total number of nucleons ($A_{\mathrm{tot}}$),
the total number of protons ($Z_{\mathrm{tot}}$), the temperature ($T$) and the volume ($V$). 
Net charge neutrality is assumed to hold locally, which means that at any point the proton density $n_p=Z_{\mathrm{tot}}/V$ is identically equal to the net density of electrons $\bar n_{\mathrm{el}}=\bar N_{\mathrm{el}}/V$.
For convenience, in the following we shall refer to the electron-positron gas as electron gas and omit the bar over $n_{\mathrm{el}}$.

The ensemble of nuclei is assumed to form an ideal gas. 
Unbound nucleons are assumed to have a homogeneous distribution. Their self-interactions are accounted for within the mean-field theory of NM, for which the non-relativistic model~\citep{Negele_PRC_1972,Vautherin1996} is employed here.  
Interactions among nuclei and with unbound nucleons occur via the excluded volume. 
As we shall see later, this is essential for the transition to homogeneous matter at densities $n_B$ of the order of $\sat{n}/2$.
The nuclear component interacts with the electron gas via electrostatic interactions only.

These assumptions allow one to decompose the total free energy of the system into the sum of free energies,
\be
F=\sum_{A,Z} (F_{A,Z}^{\mathrm{ideal}} +F_{A,Z}^{\mathrm{int}} )+F_{\mathrm{Coul}}
+F_{\mathrm{MF}}+F_{\mathrm{el}}.
\label{eq:Ftot}
\ee

The ideal-part of the free energy of a generic nucleus with the mass number $A$ and charge number $Z$ is
\ba
F_{A,Z}^{\mathrm{ideal}}(N_{A,Z}, T)&=& N_{A,Z} M_{A,Z}c^2 \nonumber \\
&-& N_{A,Z} T
\left[ \ln \left( \frac{V_{\mathrm{cl}}^{\mathrm {free}}}{N_{A,Z}} \left(\frac{M_{A,Z} T}{2 \pi \hbar^2}\right)^{3/2}\right)
  +1\right], \nonumber \\
\label{eq:FAZideal}
\ea
where $M_{A,Z} c^2$ is the rest-mass contribution and $M_{A,Z}$ and $N_{A,Z}$ denote the mass and number of the $(A,Z)$ nuclei;
$V_{\mathrm{cl}}^{\mathrm {free}}=V-V_{\mathrm {gas}}-V_{\mathrm{cl}}$ represents the volume available for the translational movement of nuclei after the volumes occupied by nucleons and all nuclear species is subtracted;
$V_{\mathrm {gas}}=(N_n+N_p) v_0$ represents the intrinsic volume of the unbound $N_{n}$ neutrons and $N_p$ protons, with each occupying the volume $v_0=1/\sat{n}$;
$V_{\mathrm{cl}}=\sum_{A,Z} N_{A,Z} v_{A,Z}$ stands for the intrinsic volume of the all nuclei.
To account for the isospin asymmetry ($\delta=1-2Z/A$) dependence of the volume, we take
$v_{A,Z}=A/\sat{n}(\delta)$.

Population of excited states, made possible by finite temperatures, asks for consideration of
the contribution from internal degrees of freedom in addition to the one due to translation.
This contribution for the species $(A,Z)$ is
\be
F_{A,Z}^{\mathrm{int}}=-T \ln z_{A,Z}^{\mathrm{int}}=-T N_{A,Z} \ln \tilde z_{A,Z}^{\mathrm{int}},
\label{eq:Fint}
\ee
depends on the species abundance and its excitation energy spectrum. 
\ba
\tilde z_{A,Z}^{\mathrm{int}}&=&\sum_j g_j \exp(-\epsilon_j/T) \nonumber \\
&=&g_{\mathrm{G.S.}}+\int_0^{B(A,Z)} d \epsilon \rho(A,Z,\epsilon) \exp(-\epsilon/T),
\label{eq:zint}
\ea
represents the internal partition function of the $(A,Z)$ nucleus; $g_j$ and $\epsilon_j$ stand for the spin degeneracy and energy of the $j$ excited state; $g_{\mathrm{G.S.}}$ is the spin degeneracy of the ground state; $\rho(A,Z,\epsilon)$ is the level density;
$B(A,Z)=(A-Z) m_n c^2 + Z m_p c^2- M_{A,Z} c^2$ is the binding energy of the nucleus. 

The Coulomb contribution to the free energy is the sum of the Coulomb energies of each nucleus and its surrounding gas of electrons,
\be
F_{\mathrm{Coul}}=\sum_{A,Z} N_{A,Z} E_{A,Z}^{\mathrm{Coul}},
\label{eq:FCoul}
\ee
where~\citep{Lattimer_NPA_1985},
\be
E_{A,Z}^{\mathrm{Coul}}=-\frac35 \frac{Z^2 e^2}{R_{A,Z}} \left( \frac32 x - \frac12 x^3\right),
\label{eq:ECoul}
\ee
with $x=\left[A/Z \cdot n_{\mathrm{el}}/\sat{n}(\delta) \right]^{1/3}$,
$R_{A,Z}=\left[3 v_{A,Z}/4\pi \right]^{1/3}$,
$n_{\mathrm{el}}=\left[N_p + \sum_{A,Z} Z N_{A,Z}\right]/V$. 
$E_{A,Z}^{\mathrm{Coul}}$ above has two components of opposite signs.
The attractive component comes from the interaction between protons and electrons. The negative one is due to the mutual repulsion between electrons.
Both terms depend on the electron density and spatial extension of the nucleus.
Notice that the electrostatic interaction among the protons in a nucleus does not appear in Eq.~\eqref{eq:ECoul}, because it is accounted for in the nuclear mass entering Eq.~\eqref{eq:FAZideal}. 

The total free energy of the unbound nucleons can be expressed as
\begin{align}
	\begin{split}
		F_{\mathrm{MF}}(N_n, N_p, T)&= N_n m_n c^2 + N_p m_p c^2 \\
		&+ V_{\mathrm{gas}}^{\mathrm{free}} f_{\mathrm{MF}}^0\left(\frac{N_n}{V_{\mathrm{gas}}^{\mathrm{free}}}, \frac{N_p}{V_{\mathrm{gas}}^{\mathrm{free}}}, T \right),
	\end{split}
	\label{eq:FMF}
\end{align}
where $V_{\mathrm{gas}}^{\mathrm{free}}=(V-V_{\mathrm{cl}})$ is the volume available for these nucleons.
As in Eq.~\eqref{eq:FAZideal}, the rest-mass contribution is explicitly written.
Notice that, at variance with the nuclei, unbound nucleons do not geometrically exclude each other. This treatment is consistent with the mean-field approach
used to model their behavior, where nucleons are  described by plane waves.
$f_{\mathrm{MF}}^0$ represents the mean-field value of the free energy density of a gas with the neutron and proton densities $\tilde n_n=N_n/V_{\mathrm{gas}}^{\mathrm{free}}$ and $\tilde n_p=N_p/V_{\mathrm{gas}}^{\mathrm{free}}$ and that does not interact with other species (thus, the superscript ``0'').
As a matter of fact, throughout this paper we mark by a ``0''  superscript the values that state variables take
when the subsystem under consideration is treated as standalone.
Examples of interactions that are already considered are:
the interactions among unbound nucleons that are accounted for in the mean-field model,
the interactions among unbound nucleons and nuclei that are modeled via the excluded volume,
the Coulomb interactions between the nuclei and the electrons, and among electrons that are explicitly dealt with in the Coulomb terms. 

The free energy of the electron gas with density $n_{\mathrm{el}}$ is
\be
F_{\mathrm{el}}(N_{\mathrm{el}},T)=V f_{\mathrm{el}}^0(n_{\mathrm{el}},T).
\ee

In the following, we shall derive the expressions for some basic thermodynamic state variables upon differentiation
of the free energy in Eq.~\eqref{eq:Ftot}.

Let us start with the chemical potential of an $(A,Z)$-nucleus defined as
\ba
\mu_{A,Z}&=&\left. \frac{\partial F}{\partial N_{A,Z}} \right|_{T,V,N_{\mathrm{el}}, N_n, N_p, \{N_{A',Z'}\}} \nonumber \\
&=& \mu_{A,Z}^{\mathrm{ideal}} + \mu_{A,Z}^{\mathrm{int}} + \mu_{A,Z}^{\mathrm{Coul}}+\mu_{A,Z}^{\mathrm{MF}}+\mu_{A,Z}^{\mathrm{el}} ,
\label{eq:definition_muAZ}
\ea
and explicitate the contribution coming from each of the terms in Eq.~\eqref{eq:Ftot}.

We notice that, in addition to an explicit dependence on $N_{A,Z}$, $F_{A,Z}^{\mathrm{ideal}}$ depends on $N_{A,Z}$ also via $V_{\mathrm{cl}}^{\mathrm{free}}$.
The first of these dependencies gives rise to
\be
\mu_{A,Z}^{\mathrm{ideal};1}=M_{A,Z}c^2-T \ln \left[ \frac{V_{\mathrm{cl}}^{\mathrm{free}}}{N_{A,Z}}
\left( \frac{M_{A,Z}T}{2 \pi \hbar^2} \right)^{3/2}
    \right] +T \frac{v_{A,Z} N_{A,Z}}{V_{\mathrm{cl}}^{\mathrm{free}}}, 
\ee
while the second leads to
\be
\mu_{A,Z}^{\mathrm{ideal};2}=T\frac{v_{A,Z}}{V_{\mathrm{cl}}^{\mathrm{free}}} \sum_{A',Z' \neq A,Z} N_{A',Z'}.
\ee
From these two contributions, one obtains
\begin{align}
	\begin{split}
		\mu_{A,Z}^{\mathrm{ideal}}=M_{A,Z}c^2 &- T \ln \left[ \frac{V_{\mathrm{cl}}^{\mathrm{free}}}{N_{A,Z}} \left( \frac{M_{A,Z}T}{2 \pi \hbar^2} \right)^{3/2} \right] \\
		&+ T \frac{v_{A,Z}}{V_{\mathrm{cl}}^{\mathrm{free}}}\sum_{A',Z'} N_{A',Z'},
	\end{split}
	\label{eq:muAZideal}
\end{align}
where the summation in the last term applies to all species, without exception.

The contributions from internal excitation and Coulomb interaction are trivial and write
\be
\mu_{A,Z}^{\mathrm{int}} = -T \ln \tilde z_{A,Z}^{\mathrm{int}},
\label{eq:muAZint}
\ee
\be
\mu_{A,Z}^{\mathrm{Coul}}=E_{A,Z}^{\mathrm{Coul}},
\label{eq:muAZCoul}
\ee
respectively.

Now, $N_{A,Z}$ enters the mean-field contribution to the total free energy via the volume of the $(A,Z)$ nuclei that
is forbidden to the unbound nucleons. As such,
\begin{align}
	&\begin{aligned}
		\mu_{A,Z}^{\mathrm{MF}}=-v_{A,Z}f_{\mathrm{MF}}^0(\tilde n_n, \tilde n_p) + (V-V_{\mathrm{cl}})
		&\left[ \frac{\partial f_{\mathrm{MF}}^0}{\partial \tilde n_n} \frac{\partial \tilde n_n}{\partial N_{A,Z}} \right. \nonumber \\
		&\left. + \frac{\partial f_{\mathrm{MF}}^0}{\partial \tilde n_p} \frac{\partial \tilde n_p}{\partial N_{A,Z}} \right] \nonumber
	\end{aligned} \\
	&=-v_{A,Z}f_{\mathrm{MF}}^0(\tilde n_n, \tilde n_p) +\mu_{\mathrm{MF};n}^0 \frac{v_{A,Z} N_n}{(V-V_{\mathrm{cl}})} +\mu_{\mathrm{MF};p}^0 \frac{v_{A,Z} N_p}{(V-V_{\mathrm{cl}})}
	\nonumber \\
	&=v_{A,Z} p_{\mathrm{MF}}^0(\tilde n_n, \tilde n_p),
  	\label{eq:muAZMF}
\end{align}
where $\mu_{\mathrm{MF};i}^0$ with $i=n,p$ represent the mean-field chemical potentials of neutrons and protons with densities $\tilde n_i$.

Considering now that the electron free energy depends only on control quantities, e.g., total number of protons, volume, and temperature, one gets $\mu_{A,Z}^{\mathrm{el}}=0$.

Putting together the expressions in Eqs.~\eqref{eq:muAZideal}, \eqref{eq:muAZint}, \eqref{eq:muAZCoul} and \eqref{eq:muAZMF},
it turns out that the chemical potential of a $(A,Z)$-nucleus is given by
\ba
\mu_{A,Z}&=&M_{A,Z}c^2-T \ln \left[ \frac{V_{\mathrm{cl}}^{\mathrm{free}}}{N_{A,Z}}
\left( \frac{M_{A,Z}T}{2 \pi \hbar^2} \right)^{3/2} \right]
+T \frac{v_{A,Z}}{V_{\mathrm{cl}}^{\mathrm{free}}}\sum_{A',Z'} N_{A',Z'} \nonumber \\
&-& T \ln \tilde z_{A,Z}^{\mathrm{int}}+E_{A,Z}^{\mathrm{Coul}}+v_{A,Z} p_{\mathrm{MF}}^0(\tilde n_n, \tilde n_p).
\label{eq:muAZ}
\ea

The chemical potential of unbound neutrons and protons can be obtained from the thermodynamic definition, 
\be
\mu_{n/p}=\left. \frac{\partial F}{\partial N_{n/p}} \right|_{T,V,N_{\mathrm{el}}, N_{p/n}, \{N_{A,Z}\}},
\ee
and leads to
\be
\mu_{n/p}=m_{n/p}c^2 + \frac{T v_0}{V_{\mathrm{cl}}^{\mathrm{free}}}\sum_{A,Z}N_{A,Z}+\mu_{\mathrm{MF};n/p}^0.
\label{eq:mugas}
\ee
The above equation shows that, due to the interaction with the gas of nuclei, the chemical potentials of the unbound nucleons are shifted with respect to the values they would have if other particles did not exist in the considered volume.

On the same line, the chemical potential of electrons,
\be
\mu_{\mathrm{el}}=\left. \frac{\partial F}{\partial N_{\mathrm{el}}} \right|_{T,V,N_n, N_p, \{N_{A,Z}\}},
\label{eq:muel}
\ee
gets modified by the Coulomb interaction with the nuclei and among themselves, which brings to the expression
\be
\mu_{\mathrm{el}}=\mu_{\mathrm{Coul}}+\mu_{\mathrm{el}}^0,
\ee
where
\be
\mu_{\mathrm{Coul}}=-\frac3{10} \sum_{A,Z} N_{A,Z} \frac{Z^2 e^2}{R_{A,Z}} \left(1-x^2\right) \frac{x}{N_{\mathrm{el}}} ,
\label{eq:muelCoul}
\ee
and $\mu_{\mathrm{el}}^0$ represents the chemical potential the electrons would have if they did not suffer any interaction.

Let us now turn to the pressure, defined as
\be
P=-\left. \frac{\partial F}{\partial V}\right|_{T, N_{\mathrm{el}}, N_n, N_p, \{N_{A,Z}\}},
\label{eq:definitionP}
\ee
which can be expressed as a sum of contributions that stem from the various terms in the free energy, see Eq.~\eqref{eq:Ftot},
\be
P=\sum_{A,Z} P_{A,Z}^{\mathrm{ideal}}+\sum_{A,Z} P_{A,Z}^{\mathrm{int}}+P_{\mathrm{Coul}}+P_{\mathrm{MF}}+P_{\mathrm{el}},
\ee
with
\be
P_{A,Z}^{\mathrm{ideal}}=T N_{A,Z} \frac1{V_{\mathrm{cl}}^{\mathrm {free}}},
\ee
\be
P_{A,Z}^{\mathrm{int}}=0,
\ee
\ba
P_{\mathrm{Coul}}&=&\frac35 \sum_{A,Z} N_{A,Z} \frac{Z^2 e^2}{R_{A,Z}} \frac32 (1 - x^2) \left(\frac{A}{Z \sat{n}(\delta)}\right)^{1/3}
\frac13 n_{\mathrm{el}}^{-2/3} \frac{\partial n_{\mathrm{el}}}{\partial V} \nonumber \\
&=& - \frac{3}{10}  \sum_{A,Z} N_{A,Z} \frac{Z^2 e^2}{R_{A,Z}} \left(1 - x^2\right) \frac{x}{V}  \nonumber \\
&=&n_{\mathrm{el}} \mu_{\mathrm{Coul}},
\label{eq:PCoul}
\ea
\begin{align}
	&\begin{aligned}
		P_{\mathrm{MF}}=-f_{\mathrm{MF}}^0(\tilde n_n, \tilde n_p)+(V-V_{\mathrm{cl}}) 
		&\left[\mu_{\mathrm{MF};n}^0 \frac{\tilde n_n}{V-V_{\mathrm{cl}}} \right. \nonumber\\
		&\left. + \mu_{\mathrm{MF};p}^0 \frac{\tilde n_p}{V-V_{\mathrm{cl}}} \right] \nonumber
	\end{aligned}\\
	& = p_{\mathrm{MF}}^0(\tilde n_n, \tilde n_p),
\end{align}
and
\be
P_{\mathrm{el}}=P_{\mathrm{el}}^0.
\ee

Similarly, the entropy of the mixture of unbound nucleons, nuclei, and electrons,
\be
S=-\left. \frac{\partial F}{\partial T}\right|_{V, N_{\mathrm{el}}, N_n, N_p, \{N_{A,Z}\}},
\label{eq:definitionS}
\ee
can be decomposed into
\be
S=\sum_{A,Z} S_{A,Z}^{\mathrm{ideal}}+\sum_{A,Z} S_{A,Z}^{\mathrm{int}}+S_{\mathrm{Coul}}+S_{\mathrm{MF}}+S_{\mathrm{el}},
\ee
where
\be
S_{A,Z}^{\mathrm{ideal}}=N_{A,Z} \left[ \ln \left(\frac{V_{\mathrm{cl}}^{\mathrm {free}}}{N_{A,Z}} \left(\frac{M_{A,Z} T}{2 \pi \hbar^2} \right)^{3/2}\right) + \frac52 \right], 
\ee
\be
S_{A,Z}^{\mathrm{int}}=N_{A,Z} \left[ \ln \tilde z_{A,Z}^{\mathrm{int}} + T \frac{\partial \ln \tilde z_{A,Z}^{\mathrm{int}}}{\partial T} \right], 
\ee
\be
S_{\mathrm{Coul}}=0,
\ee
\be
S_{\mathrm{MF}}=S_{\mathrm{MF}}^0 (\tilde n_n, \tilde n_p),
\ee
\be
S_{\mathrm{el}}=S_{\mathrm{en}}^0 (N_{\mathrm{en}}^0).
\ee

The total energy can be computed from the thermodynamic identity,
\be
E=F+TS,
\ee
and provides,
\ba
E&=&\sum_{A,Z} N_{A,Z} \left[M_{A,Z} c^2 +
  \frac32 T + \langle \epsilon_{A,Z} \rangle  + E_{A,Z}^{\mathrm{Coul}} \right] \nonumber \\
&+& (V-V_{\mathrm {cl}}) \left[m_n c^2 \tilde n_n +m_p c^2 \tilde n_p + e_{\mathrm{MF}}^0(\tilde n_n, \tilde n_p) \right]\nonumber \\
&+&V e_{\mathrm{el}}^0(n_{\mathrm{el}}),
\label{eq:Etot}
\ea
where
\ba
\langle \epsilon_{A,Z} \rangle&=&\frac{\int_0^{B(A,Z)} d \epsilon \epsilon \rho(\epsilon) \exp(-\epsilon/T) }
{\int_0^{B(A,Z)} d \epsilon \rho(\epsilon) \exp(-\epsilon/T)}
\nonumber \\
&=& T^2  \frac{\partial \ln \tilde z_{A,Z}^{\mathrm{int}}}{\partial T},
\ea
represents the average excitation energy of the nucleus $(A,Z)$.

It is interesting to point out that the thermodynamic identity $F-PV+\sum_{i=n,p,\{A,Z\},\mathrm{el}} \mu_i N_i=0$ holds for the mixture as a whole but not for individual components, that is, $F_i -P_i V -\mu_i N_i \neq 0$ where $i$ corresponds to each of the following: the gas of unbound self-interacting nucleons, the ensemble of nuclei and the electron gas.
This situation is due to the strong coupling between the three sub-systems and the fact that it is impossible to attribute the Coulomb terms to any particular one.

Strong and electromagnetic equilibrium between the homogeneous and clusterized nuclear components requires that
\be
\mu_{A,Z}=(A-Z) \mu_n + Z \mu_p.
\ee
Substitution of expressions in Eqs.~\eqref{eq:muAZ} and \eqref{eq:mugas} leads to
\ba
-T \ln \left[\frac{V_{\mathrm{cl}}^{\mathrm{free}}}{N_{A,Z}} \left( \frac{M_{A,Z}T}{2 \pi \hbar^2}\right)^{3/2} \tilde z_{A,Z}^{\mathrm{int}}\right]&=&
(A-Z) \mu_{\mathrm{MF}, n}^0 +Z \mu_{\mathrm{MF},p}^0 \nonumber \\
& + & (A-Z) m_n c^2 + Z m_p c^2 \nonumber \\
& - & E_{A,Z}^{\mathrm{Coul}}-A v_0 p_{\mathrm{MF}}^0 - M_{A,Z} c^2 , \nonumber \\
\ea
from which it is straightforward to extract the number of $(A,Z)$-nuclei in the volume $V$,
\begin{align}
&N_{A,Z} = V_{\mathrm{cl}}^{\mathrm{free}} z_{A,Z}^{\mathrm{int}} \left( \frac{M_{A,Z}T}{2 \pi \hbar^2}\right)^{3/2} \nonumber \\
&\times \exp \left(\frac{(A-Z) \mu_{\mathrm{MF}, n}^0 +Z \mu_{\mathrm{MF},p}^0 - E_{A,Z}^{\mathrm{Coul}}-A v_0 p_{\mathrm{MF}}^0 +B(A,Z)}{T} \right).
\label{eq:NAZ}
\end{align}

Equation~\eqref{eq:NAZ} reveals that an essential piece of information for determining the
composition of the inhomogeneous phase is represented by the mean-field values of the chemical potentials of the unbound self-interacting nucleons. For fixed temperature $T$,
density $n_B$ and proton fraction $Y_p$, these are determined by solving the equations for
mass and charge number conservation,
\ba
n_B V=A_{\mathrm{tot}}&=&N_n+N_p+\sum_{A,Z} N_{A,Z} A,\\
n_B Y_p V=Z_{\mathrm{tot}}&=&N_p+\sum_{A,Z} N_{A,Z} Z.
\label{eq:conserveq}
\ea
Despite its simplicity, the numerical resolution of the above system is highly non-trivial.
The reason is that, via $V_{\mathrm{cl}}^{\mathrm{free}}$, $N_{A,Z}$ depends on itself and on the intrinsic volume of unbound nucleons, which, in turn, depends on the unknown variables $\mu_{\mathrm{MF}, n/p}^0$.
We found it convenient to introduce a third unknown variable, $V_{\mathrm{cl}}$, which makes the computation of $V_{\mathrm{cl}}^{\mathrm{free}}$ in Eq.~\eqref{eq:NAZ} straightforward, and constrain its value by a third equation,
\be
V'_{\mathrm {cl}}=\sum_{A,Z} N_{A,Z} v_{A,Z}.
\label{eq:consistVcl}
\ee

Once $\mu_{\mathrm{MF}, n/p}^0$ are known, both the composition of matter and its energetics can be determined using the equations in this section.

A word of caution is in order here.
The definition of the chemical potential of an $(A,Z)$-nucleus in Eq.~\eqref{eq:definition_muAZ}, which is a key ingredient of our derivation, assumes that variations of the total free energy with respect to the number of particles belonging to one species can be taken without altering the rest of the cluster gas composition.
This assumption is consistent with the ideal gas working frame, where no interaction exists among the nucleus under consideration and other nuclei that form the gas of nuclear clusters.
Similarly, Eqs.~\eqref{eq:definitionP} and \eqref{eq:definitionS} assume that derivatives of the total free energy with respect to volume and temperature can be taken without modifying the composition of the mixture.
Yet, Eq.~\eqref{eq:NAZ} shows that $\{N_{A,Z}\}$ are strongly coupled and the coupling becomes more important at high densities;
it also shows that $\{N_{A,Z}\}$ has an explicit dependence on temperature.
Implicit dependence also exists through all ingredients other than $M_{A,Z}$ and $B(A,Z)$.
Note that all these dependencies arise exclusively due to the excluded volume.
This corroborates previous microscopic evidence~\citep{Typel_EPJA_2016} that, in spite of its apparent simplicity, the excluded volume approximation is very efficient in effectively accounting for interactions.

All these equations are similar to those of \cite{Hempel_NPA_2010}. Nevertheless, the derivation is different.

\subsection{Nuclei}
\label{ssec:Nuclei}

In this work, we allow all nuclei with $A \geq 2$, for which experimental data and/or theoretical evaluations of the mass are available, to be populated in the inhomogeneous phase. 
For experimental values of nuclear masses, we use the AME2020~\citep{AME2020} table, which is the most recent available to date.
For theoretical estimations, we use the 10 parameter mass table by \cite{Duflo_PRC_1995}\,\footnote{The table is taken from \href{https://www-nds.iaea.org/amdc/theory/du_zu_10.feb96}{https://www-nds.iaea.org/amdc/theory/du\_zu\_10.feb96}}.
The heaviest nuclei present in \citep{AME2020} and \citep{Duflo_PRC_1995} have $A=295$ and 297, respectively.
The advantage of supplementing the pool of nuclei in the experimental mass table with nuclei for which mass estimates have been performed with dedicated theoretical models consists in accounting for nuclei far beyond the drip lines, which are expected to exist in astrophysical environments like those produced at different stages in the evolution of supernovae or binary mergers. 
With peculiar reaction rates, exotic nuclei impact the dynamical evolution as well as the nucleosynthesis of heavy nuclei at astrophysical sites, where they serve as initial conditions for nuclear reaction network calculations. 

The isospin dependence of the nuclear volumes is implemented via the isospin dependence of the saturation density of NM, $v_{A,Z}=A/\sat{n}(\delta)$, as already explained in the previous section. 
For nuclei with an isospin asymmetry larger than the maximum value ($\delta_{\mathrm{max}}$) for which NM manifests saturation, $v_{A,Z}=A/\sat{n}(\delta_{\mathrm{max}})$.

The ground states of even (odd) mass nuclei are considered to have a spin degeneracy factor $g_{\mathrm{G.S.}}=1~(2)$.
For nuclei with $A \geq 16$, the temperature-triggered population of excited states is accounted for via the energy-dependent back-shifted Fermi gas level density; in particular, we adopt the parametrization proposed by \cite{Bucurescu_PRC_2005,vonEgidy_PRC_2005}.
The upper limit of the excitation energy spectrum in the expression for the internal partition function of an $(A,Z)$-species is taken here to be $B(A,Z)$, see Eq.~\eqref{eq:zint}.

The presented modeling of nuclear clusters in a hot and dense medium suffers from two obvious drawbacks.
First, in-medium modifications of the nuclear surface energy, which increase with the density of the surrounding medium and are essential for the dissolution of nuclei in dense nucleonic environments, are treated phenomenologically.
A more accurate implementation of such effects requires dedicated microscopic studies, which are not available yet.
Second, by disregarding temperature effects on the binding energy, our approach allows nuclei to survive beyond the limiting temperature of the Coulomb instability~\citep{Jaqaman_PRC_1984,Song_PRC_1991}.
This feature is common for many eNSE implementations~\citep{Hempel_NPA_2010,Furusawa_ApJ_2011,Furusawa_ApJ_2013,Furusawa_NPA_2017,Raduta_NPA_2019} and explains why a significant fraction of nuclei with various masses exists at $T>15~\mathrm{MeV}$. For a comparison, see Fig.~17 in \citep{Raduta_EPJA_2021}. 

To correct for the spurious persistence of nuclei at temperatures higher than 8-12 MeV,
in our model inhomogeneous matter with sub-saturation densities and  $T \geq 2 T_C/3$, where $T_C$ stands for the critical temperature of the liquid-gas phase transition of sub-saturated homogeneous matter, is superseded by homogeneous matter.
The value we choose for the limiting temperature of the Coulomb instability is based on microscopic calculations of \cite{Jaqaman_PRC_1984,Song_PRC_1991} and, in contrast with the results of \cite{Jaqaman_PRC_1984,Song_PRC_1991}, disregards extra finite size and isospin dependencies.

\subsection{Nucleons and effective interactions}
\label{ssec:Nucleons}

\renewcommand{\arraystretch}{1.0}
\setlength{\tabcolsep}{3.9pt}
\begin{table*}
  \caption{Parameters of the five Brussels extended Skyrme effective interactions for which general-purpose EOS tables are built in this work.
    For all these forces, $x_4=x_5=0$.
    BBSk1 is extracted from the run~2 of \citep{Beznogov_PRC_2024} and represents our benchmark model;
    BBSk2, BBSk3, BBSk4 and BBSk5 are extracted from run~1 of \citep{Beznogov_PRC_2024}. 
  }
  \label{tab:effintparam}
  \centering
	\begin{tabular}{ccrrrrr}
	  \toprule
	  \toprule
          Param. & Units & BBSk1 & BBSk2 & BBSk3 & BBSk4 & BBSk5 \\
          \midrule
          $C_0$   & MeV fm$^3$ & $-554.2673$   & $-473.5086$ & $-576.5550$ & $-528.1041$ & $-843.1804$ \\
          $D_0$   & MeV fm$^3$ & $ 246.2854$   & $ 220.3469$ & $ 262.5667$ & $ 260.5593$ & $ 334.0308$ \\
          $C_3$   & MeV fm$^{3+3 \sigma}$ & $ 687.6303$   & $ 715.1548$ & $ 641.6009$ & $ 768.6374$ & $ 907.4757$ \\
          $D_3$   & MeV fm$^{3+3 \sigma}$ & $-234.5704$   & $-242.6774$ & $-239.1944$ & $-162.6620$ & $-302.8400$ \\
          $\eff{C}$ & MeV fm$^5$ &  $538.0267$ & $ 38.86668$ & $ 793.6469$ & $ 667.9141$ &  8.696107 \\
          $\eff{D}$ & MeV fm$^5$ & $-234.5062$ & $-31.28217$ & $-345.4996$ & $-478.3312$ & $0.700793$ \\
          $t_4$     & MeV fm$^{5+3 \beta}$  & $-3509.407$ & $-454.9495$ & $-5096.015$ & $-6388.676$ & $ 225.8386$ \\
          $t_5$     & MeV fm$^{5+3 \gamma}$ & $ 393.0868$ & $ 132.9851$ & $ 553.4818$ & $ 1606.935$ & $-71.20556$ \\
          $\sigma$   &  ---         & 0.581786 & 0.681196 & 0.695421 & 1.084970 & 0.228234 \\
          $\beta$    &  ---         & 0.136059 & 0.936844 & 0.168920 & 0.196900 & 0.694620 \\
          $\gamma$   &  ---         & 0.483806 & 1.043440 & 0.607837 & 0.209448 & 0.335074 \\     
          \bottomrule
	  \bottomrule
	\end{tabular}
\end{table*}
\renewcommand{\arraystretch}{1.0}
\setlength{\tabcolsep}{2.0pt}

\renewcommand{\arraystretch}{1.0}
\setlength{\tabcolsep}{3.7pt}
\begin{table}  
  \caption{Values of selected NM parameters corresponding to the effective interactions in Table~\ref{tab:effintparam}.
    Provided are: the saturation density ($\sat{n}$) of SNM; energy per nucleon ($\sat{E}$), compression modulus ($\sat{K}$),
    skewness ($\sat{Q}$) and kurtosis ($\sat{Z}$) of SNM; the critical temperature ($T_C$) for the liquid-gas phase transition of SNM;
    the symmetry energy ($\sym{J}$), its slope ($\sym{L}$),
    compressibility ($\sym{K}$), skewness ($\sym{Q}$) and kurtosis ($\sym{Z}$); Landau effective mass of the nucleons
    in SNM ($m_{\mathrm{eff;\,N}}^\mathrm{SNM}$) and Landau effective mass of the neutrons in PNM ($m_{\mathrm{eff;\,n}}^\mathrm{PNM}$) at 0.16~fm$^{-3}$.
    }
  	\label{tab:NM}
        \centering
	\begin{tabular}{ccccccc}
		\toprule
		\toprule
        Param.                               & Units                  &  BBSk1 & BBSk2 & BBSk3 & BBSk4 & BBSk5 \\
        \midrule
        $n_\mathrm{sat}$                    & $\mathrm{fm}^{-3}$    &  0.161 & 0.154 & 0.165 & 0.164 & 0.158   \\
        $E_\mathrm{sat}$                    & $\mathrm{MeV}$       & $-15.75$ & $-15.78$ & $-15.98$  & $-15.71$ & $-15.78$ \\
        $K_\mathrm{sat}$                    & $\mathrm{MeV}$       & 243   & 287 & 225 & 248 & 228  \\
        $Q_\mathrm{sat}$                    & $\mathrm{MeV}$       & $-378$ & $-271$ & $-452$ & $-369$ & $-327 $ \\
        $Z_\mathrm{sat}$                    & $\mathrm{MeV}$       & 1486  & 362 & 2084 & 1364 & 1633 \\
        $J_\mathrm{sym}$                    & $\mathrm{MeV}$       & 30.9 & 31.4 & 29.6  & 30.1 & 32.8   \\
        $L_\mathrm{sym}$                    & $\mathrm{MeV}$       & 52.6 & 55.4 & 39.6 & 56.8 & 61.3 \\
        $K_\mathrm{sym}$                    & $\mathrm{MeV}$       & $-118$ & $-146$ & $-146$ & $-55$ & $-126$ \\
        $Q_\mathrm{sym}$                    & $\mathrm{MeV}$       & 471 & 323 & 649 & 541 & 270 \\
        $Z_\mathrm{sym}$                    & $\mathrm{MeV}$       & $-2103$  & $-1339$  & $-2526$ & $-2677$ & $-1615$ \\
        $m_{\mathrm{eff;\,N}}^\mathrm{SNM}$    & $m_{\mathrm{N}}$       & 0.630 & 0.818  & 0.461 & 0.422  & 0.940  \\
        $m_{\mathrm{eff;\,n}}^\mathrm{PNM}$    & $m_{\mathrm{n}}$       & 0.852 & 0.965 & 0.725 & 0.747 & 0.979 \\
        $T_{\mathrm{C}}$                   & $\mathrm{MeV}$         & 17.21 & 17.28 & 18.23 & 19.38 & 15.84 \\
\bottomrule
		\bottomrule
	\end{tabular}
\end{table}
\renewcommand{\arraystretch}{1.0}
\setlength{\tabcolsep}{2.0pt}
  
Unbound nucleons are treated within the non-relativistic mean-field model~\citep{Negele_PRC_1972,Vautherin1996} with zero-range interactions, which successfully describes the properties of ground-state and excited nuclei and NM at both zero and finite temperatures.

In the absence of spin polarization, the energy density of homogeneous matter with no Coulomb interaction can be expressed as the sum of four terms,
\begin{equation}
  {\cal H}=k+h_0+h_3+\eff{h},
  \label{eq:h}
\end{equation}
where $k=\hbar^2 \tau/ 2 m$ is the kinetic energy term, $h_0$ is a density-independent two-body term, $h_3$ is a density-dependent term, and $\eff{h}$ is a momentum-dependent term. Each of the interaction terms can be expressed analytically in terms of densities of particles and kinetic energies, and parameters of the effective interaction.

The effective interactions used here belong to the family of Brussels extended Skyrme interactions~\citep{BSk22-BSk26}, situation in which the
density-independent two-body term and the density-dependent term have expressions identical to those obtained in the case of standard Skyrme interactions~\citep{Ducoin_NPA_2006},
\begin{align}
  &h_0 = C_0 n^2+D_0 n_3^2,
  \label{eq:h0} \\
  &h_3 = C_3 n^{\sigma+2}+D_3 n^{\sigma} n_3^2.
  \label{eq:h3} 
\end{align}
The situation of the momentum-dependent term is different. While keeping the structure it has for standard interactions,
\be
  \eff{h} = \eff{\widetilde C}(n) n \tau+\eff{\widetilde D}(n) n_3 \tau_3,
\label{eq:heff}
\ee
this term features a more complex density dependence due to the density dependence of $\eff{\widetilde C}(n)$ and $\eff{\widetilde D}(n)$. The latter relate
to the $\eff{C}$ and $\eff{D}$ parameters of the momentum-dependent terms of the standard
interactions~\citep{Ducoin_NPA_2006} via~\citep{Beznogov_PRC_2024}
\begin{align}
\begin{split}
    &\eff{\widetilde C}(n) = \eff{C} + \left[ 3 t_4 n^{\beta} +t_5 \left( 4 x_5+5\right) n^{\gamma}\right]/16, \\
    &\eff{\widetilde D}(n) = \eff{D} +\left[ -t_4 \left(2 x_4+1 \right) n^{\beta} +t_5  \left( 2 x_5+1\right) n^{\gamma}\right]/16.
  \end{split}
  \label{eq:CeffDeff_B}
\end{align}
In the equations above, $n=n_n+n_p$ and $n_3=n_n-n_p$ stand for the isoscalar and isovector particle number densities; $\tau=\tau_{n}+\tau_{p}$ and $\tau_3=\tau_{n}-\tau_{p}$ denote the isoscalar and isovector densities of kinetic energy; $2/m=1/m_{n}+1/m_{p}$, where $m_i$ with $i={n},{p}$ denotes the bare mass of nucleons. 

Our preference for Brussels extended interactions is due to the extra density-dependent terms of the momentum interaction, which makes it possible to heal a series of drawbacks commonly associated with the standard Skyrme interactions. 
In the first place, the increased flexibility of the effective Hamiltonian can be exploited to make the effective mass of the nucleons
\be
\frac{1}{{\eff{m}}_{;i}}=\frac{1}{m_i}+\frac{2}{\hbar^2} \left[\eff{\widetilde C}(n) n \pm \eff{\widetilde D}(n) n_3 \right],
\label{eq:meff}
\ee
have a density dependence similar to the one put forward by {\em ab initio} models with three body forces~\citep{Burgio_PRC_2020,Drischler_PRC_2021}.
In the limit of zero temperature, ${\eff{m}}_{;i}$ corresponds to the Landau effective mass defined in terms of density of
single-particle states at the Fermi surface,
\be
\frac{1}{{\eff{m}}_{;i}}= \left . \frac{1}{k_i} \frac{de_i}{dk_i} \right|_{k=k_{F;i}},
\ee
where $k_{F;i}$ stands for the Fermi momentum of the $i$-particle.
More precisely, instead of continuously decreasing with density, as is the case of the standard Skyrme interactions, extended interactions allow ${\eff{m}}_{;i}$ to increase with density for densities exceeding a certain value.
The bonus of banning ${\eff{m}}_{;i}$ from excessively decreasing with density is that the Fermi velocity of nucleons
\be
v_{F;i}=\frac{k_{F;i}}{{\eff{m}}_{;i}},
\label{eq:vF}
\ee
in dense matter does not exceed the speed of light, which would make the interaction unphysical~\citep{Urban_PRC_2023}. 

The five effective interactions for which we build here general-purpose EOS tables were extracted from the bunches of several hundred thousand interactions in runs~2 (BBSk1) and 1 (BBSk2 to BBSk5) in ~\citep{Beznogov_PRC_2024}, built within a Bayesian inference of the EOS of dense matter.
As every other effective interaction in \citep{Beznogov_PRC_2024}, they comply by construction with standard constraints on:
i) the best known parameters of NM,
ii) {\em ab initio} calculations of energy per particle in pure neutron matter (PNM), $(E/A)_\mathrm{PNM}$, with densities lower than $\sat{n}$, as computed by \cite{Drischler_PRC_2021}
iii) a $2~\Msun$ lower limit on the maximum gravitational mass of NS,
iv) thermodynamic stability of NS EOS,
v) causality of NS EOS up to the density corresponding to the central density of the maximum mass configuration
as well as a number of extra constraints like:
vi) for $n \leq 0.8~\mathrm{fm}^{-3}$, the Landau effective mass $\eff{m}$ of neutrons in PNM and of nucleons in symmetric NM (SNM) take values in the domain $[0, m_N]$, where $m_N$ is the bare mass,
v) for $n \leq 0.8~\mathrm{fm}^{-3}$, the neutron Fermi velocity $v_{F;n}$ in both SNM and PNM does not exceed the speed of light.
For BBSk1, which is our benchmark interaction, the effective mass of neutrons in PNM and the effective mass of nucleons in SMN also comply with {\em ab initio} calculations over $0.08 \leq n \leq 0.16~\mathrm{fm}^{-3}$.
Similarly to the values of $(E/A)_\mathrm{PNM}$, we use $\eff{m}$ predictions by \cite{Drischler_PRC_2021}.
As PNM is stiffer than NS matter, we make sure that, for the forces selected in this work, PNM is causal at least up to $0.8~\mathrm{fm}^{-3}$. 
This set of interactions was selected in such a way as to manifest the upmost variability in the behaviors of effective masses and thermal pressure as functions of density. 
BBSk1 corresponds to the ``mean'' model of run~2; BBSk2 and BBSk3 lie close to the 0.98 upper and 0.02 lower quantiles of $\eff{m}(n)$ in run~1; BBSk4 and BBSk5 have extreme behaviors of thermal pressure.

The parameters of these interactions are provided in Table~\ref{tab:effintparam} while some of their nuclear properties are listed in Table~\ref{tab:NM} in terms of NM parameters. Also given in Table~\ref{tab:NM} are the values of Landau effective mass of nucleons in SNM and Landau effective mass of neutrons in PNM at $n=0.16~\mathrm{fm}^{-3}$ as well as the values of the critical temperature of the liquid-gas phase transition in SNM. 

\section{Inhomogeneous to homogeneous matter transition}
\label{sec:Transition}

General-purpose EOS tables on \textsc{CompOSE} are tabulated as a function of temperature ($T$), baryonic charge fraction, here equal to the proton fraction ($Y_p$), and number density ($n_B$).
This format commands to transition from inhomogeneous to homogeneous matter for every value of the grid temperature lower than the value of the Coulomb instability temperature of the nuclei~\citep{Jaqaman_PRC_1984,Song_PRC_1991} and $Y_p$.

\begin{figure}
  \centering
  \includegraphics[scale=0.29]{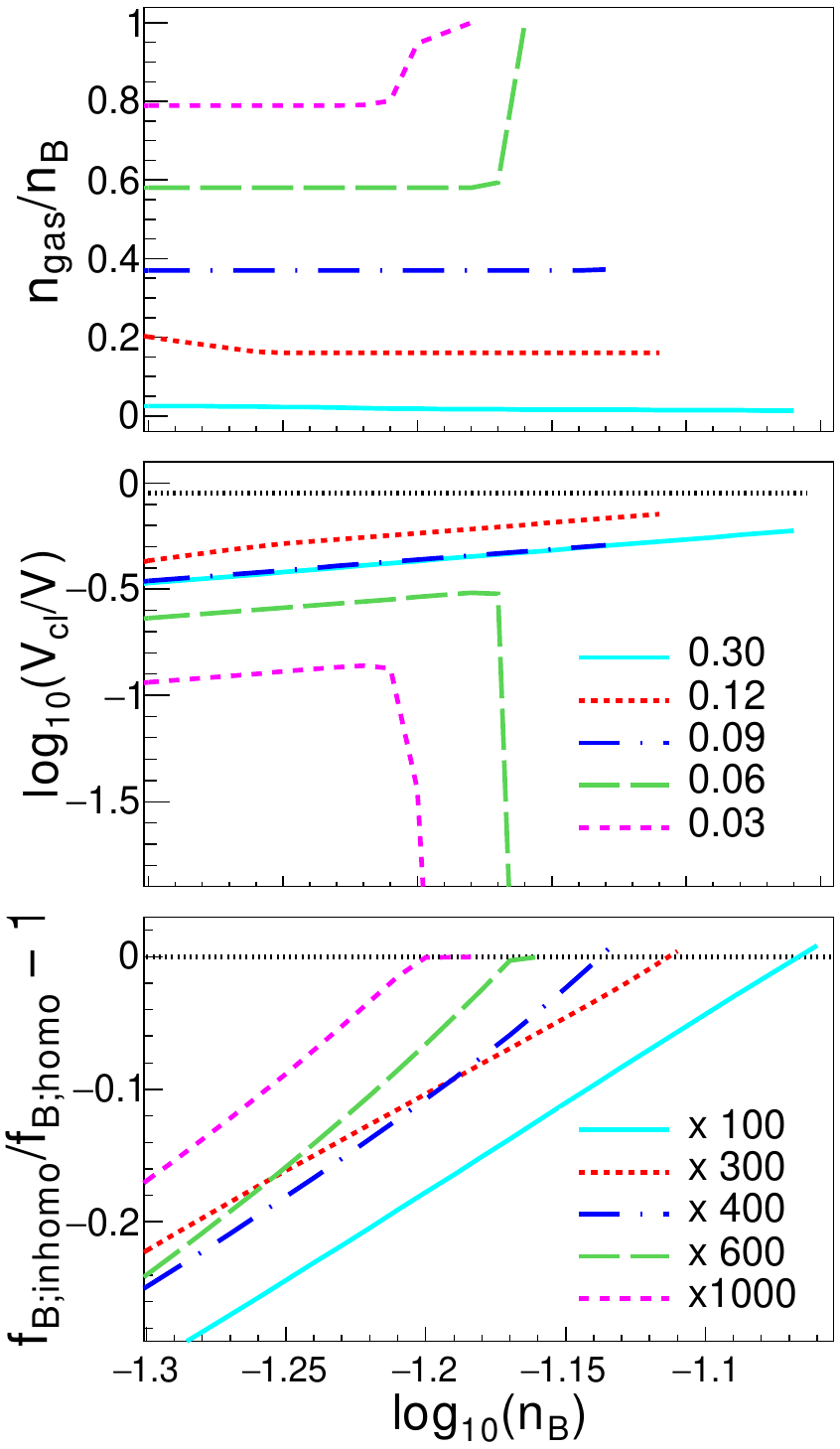}
  \includegraphics[scale=0.29]{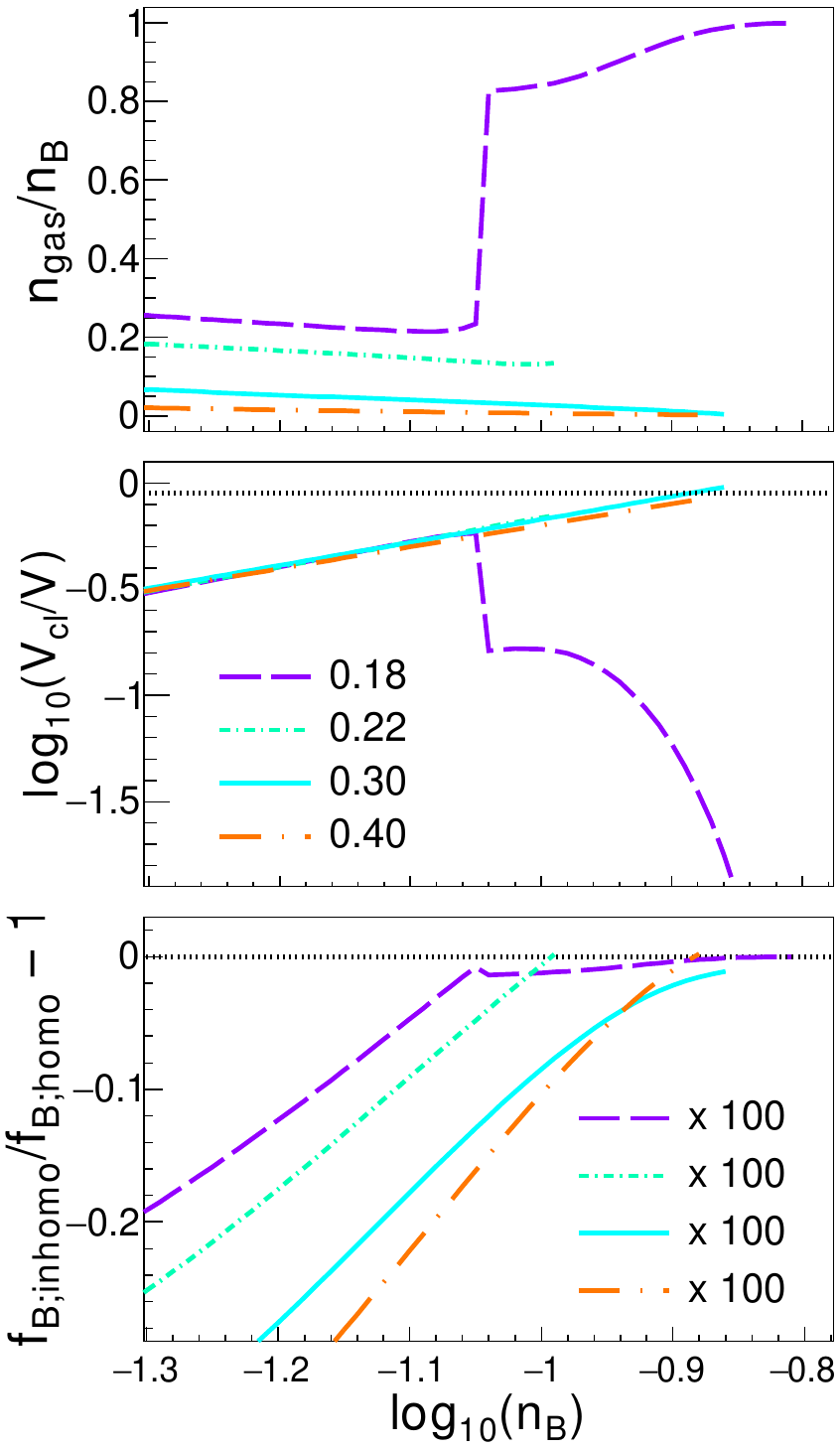}
  \caption{Selected properties of the mixture of unbound nucleons and nuclei while approaching the transition to homogeneous matter for $T=0.23~\mathrm{MeV}$ (left column) and 5.75~MeV (right column) and various proton fractions, as indicated in the legend.
  Considered are: the relative difference between the baryonic free energy densities of inhomogeneous and homogeneous matter (bottom row),
  intrinsic volume of nuclei with $A \geq 2$ relative to the full volume $V$ (middle row), and
  relative density of unbound nucleons (top row) as functions of density.
  For better readability, the quantities in the bottom panel are multiplied with arbitrary factors shown in the legend. The considered interaction is BBSk1.}
  \label{Fig:TransMech}
\end{figure}

\begin{figure}
  \centering
   \includegraphics[scale=0.4]{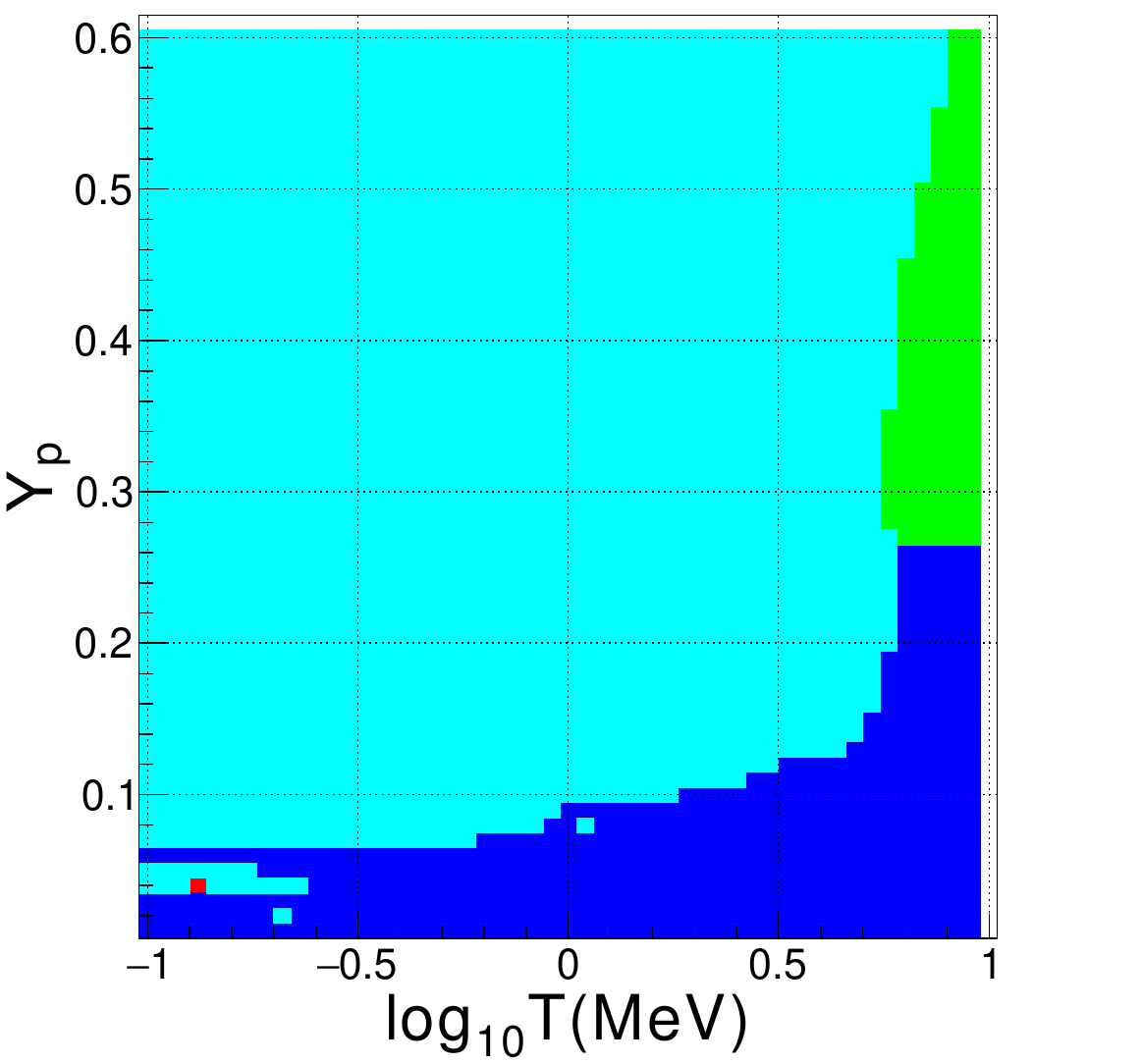}
  \includegraphics[scale=0.4]{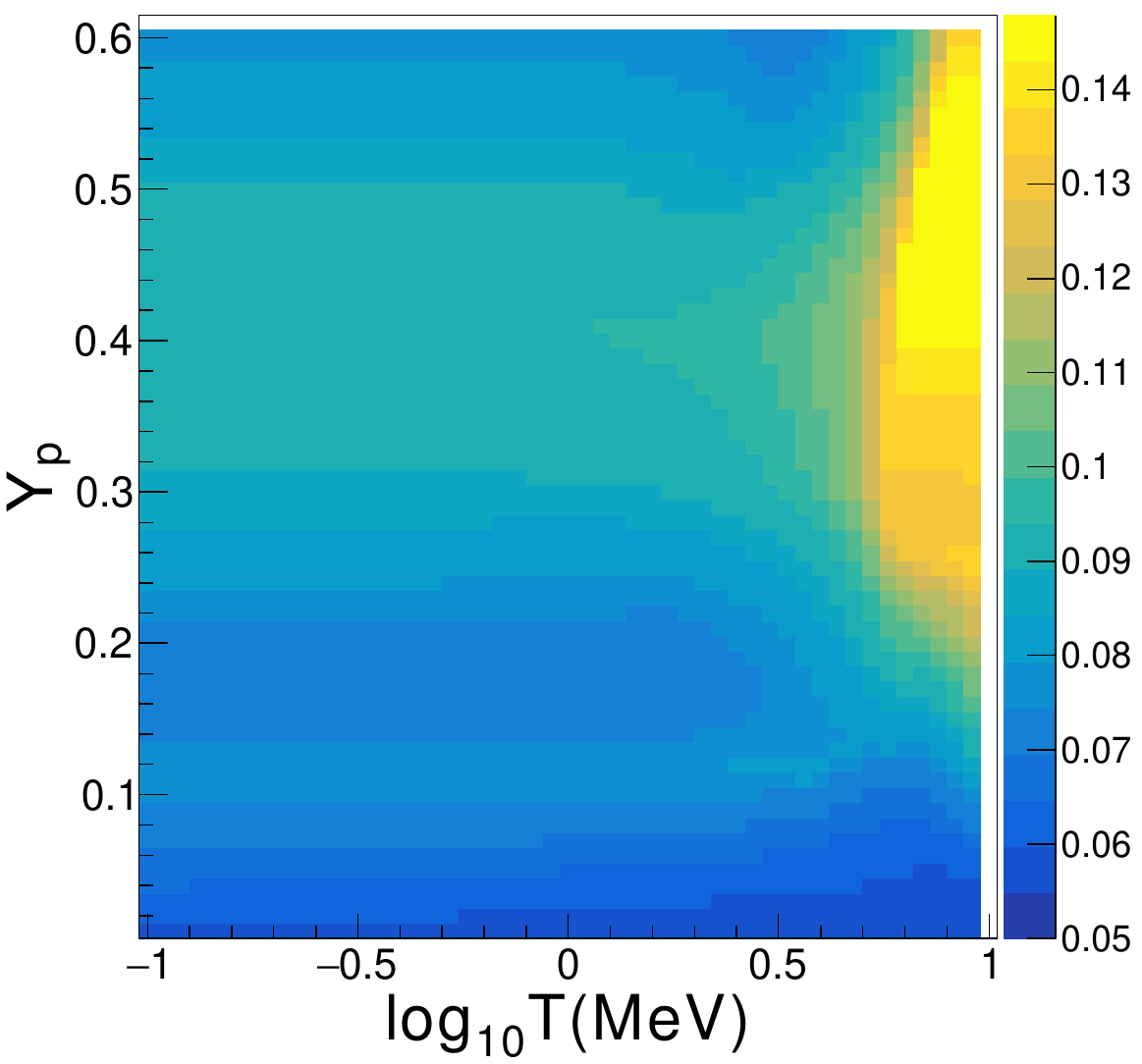}
  \caption{Mechanisms (top panel) and the density (bottom panel) for the transition to homogeneous matter as a function of temperature and proton number.
    The color legend in the top panel is as follows:
    (I) blue: $V_{\mathrm{cl}}(n)$ for fixed values of $T$ and $Y_p$ reaches the maximum;
    (II) green: $V_{\mathrm{cl}}/V \geq 0.9$;
    (III) cyan: $f_{\mathrm{inhomo}}/f_{\mathrm{homo}}-1 \geq -10^{-6}$;
    red: transition occurs at the highest density where convergence of Eqs.~\eqref{eq:conserveq} and \eqref{eq:consistVcl} is reached.
    In the bottom panel the color levels correspond to $n_{\mathrm{tr}}$ in $\mathrm{fm}^{-3}$.
    The considered interaction is BBSk1.
    \label{Fig:TransMap}
  }
\end{figure}

Implementation of nucleus-nucleus and nucleus-nucleon interactions via the excluded volume suggests that this transition arises because either (I) the uniform gas of unbound nucleons occupies the entire volume or (II) the gas of clusters occupies the entire volume.
A third mechanism consisting of a sudden replacement of the inhomogeneous phase by the homogeneous one (III) may be foreseen whenever the homogeneous phase becomes energetically favored, which corresponds to $f_{\mathrm{homo}} < f_{\mathrm{inhomo}}$ where $f$ stands for the free energy density.

Figure~\ref{Fig:TransMech} investigates the behavior of three quantities, representative for each of the three transition mechanisms, as functions of density. These are: the fraction of mass in the unbound nucleon gas, the relative volume occupied by nuclei, and the relative difference between the baryonic free energy densities of inhomogeneous and homogeneous matter. For better readability, the contributions of electrons and photons, which depend only on $n_{\mathrm{el}}$ and $T$, are removed from the free energy density.

The case of low temperatures is addressed in the left panels, where $T=0.23~\mathrm{MeV}$ and $0.03 \leq Y_p \leq 0.3$. 
It comes out that for $Y_p$ values exceeding a certain threshold, nuclear clusters bind a considerable fraction of matter. 
The increase in density results in larger fractions of volume being occupied by clusters. 
Nonetheless, the transition to homogeneous matter occurs before the clusters fill the entire volume, due to $f_{\mathrm{homo}}<f_{\mathrm{inhomo}}$. 
For lower values of $Y_p$, nuclear clusters bind a lesser fraction of matter. 
When the density exceeds a certain $Y_p$-dependent threshold, the gas of unbound nucleons suddenly becomes more dense, which forces a sudden diminish of the volume occupied by the clusters and of the mass fraction bound in the clusters.
The sudden change in the density of the gas of unbound nucleons can be explained by the jump from the low-density stable branch of  $\left.\mu_{p/n}(n_{p/n})\right|_{\mu_{n/p}}$ curve(s) to either the intermediate-density unstable branch or the high-density stable branch. We remind at this point that for $T \leq T_C$, where $T_C=15-20~\mathrm{MeV}$ is the critical temperature of the liquid-gas phase transition, uncharged homogeneous matter exhibits thermodynamic instabilities with respect to density fluctuations, which are manifested as decreasing regions of the $\mu_{p/n}(n_{p/n})$ curves at constant values of $\mu_{n/p}$~\citep{Ducoin_NPA_2006}. Consequently, multiple solutions for $(n_n$,$n_p)$ are possible for given $(\mu_n$, $\mu_p)$-sets that correspond to the phase instability window.
The residual Coulomb interaction between the relatively few and light clusters 
in the clusterized phase is important enough to prevent
$f_{\mathrm{inhomo}}$ from exceeding $f_{\mathrm{homo}}$ and, thus, be superseded by homogeneous matter. Indeed, for $Y_p=0.03$ and 0.06 and high densities, $f_{\mathrm{inhomo}} / f_{\mathrm{homo}}-1$ is almost constant and slightly lower than 0.
It is clear that, under the conditions in the left panels, only mechanisms I and III are at work. 

The right panels investigate what happens at higher temperatures, in particular $T=5.75~\mathrm{MeV}$. As before, several values of $Y_p$ are considered.
For the lowest considered value of proton fraction, $Y_p=0.18$, the situation is similar to what we observed for the two lowest values of $Y_p$ in the case of low $T$, that is, the transition occurs because the unbound nucleons kick out the clusters (I).  
For the second low value of $Y_p$, $Y_p=0.22$, the transition occurs because homogeneous matter becomes energetically favored (III).
For $Y_p=0.3$, the transition occurs because nuclear clusters occupy the entire volume (II).
For the highest considered $Y_p$ the transition occurs again because homogeneous matter is energetically more favored (III). 

A synoptic outline of the transition mechanisms as a function of $Y_p$ and $T$ is offered in Fig.~\ref{Fig:TransMap} (top panel)
for $0.01 \leq Y_p \leq 0.6$ and $T\leq 10~\mathrm{MeV}$.
The criteria upon which the transitions via mechanisms~II and III are considered to take place are $V_{\mathrm{cl}}/V \geq 0.9$
and $f_{\mathrm{inhomo}}/f_{\mathrm{homo}}-1 \geq -10^{-6}$, respectively; the value of $-10^{-6}$ arises from the precision with which the values of the free energy density of inhomogeneous and homogeneous matter are saved into the files.
The difficulty in achieving convergence of the eNSE equations when the gas of strongly interacting unbound nucleons is unstable or very dense made it preferable to consider that the transition via the mechanism~I occurs at the density where $\left.V_{\mathrm{cl}}(n)\right|_{Y_p}$ reaches the maximum.
Figure~\ref{Fig:TransMap} shows that, for low values of $Y_p$, matter transitions via mechanism I; for $5 \lesssim T \lesssim 10~\mathrm{MeV}$ and $Y_p \gtrsim 0.25$, matter transitions via mechanism II; in all other cases, the transition is done by supersession (III).
Some arbitrariness of the transition criteria is reflected in narrow domains bearing the color of mechanism III in a region where the transition holds via mechanism~I. 
There is a single set of $(T,Y_p)$ (marked in red) where difficulties in solving the eNSE equations made it impossible to reach a density high enough to see one of our transition mechanisms in operation. In this situation, the transition is done after the last converged density point.  

The transition densities are indicated in the bottom panel in Fig.~\ref{Fig:TransMap}. 
The same domains of $Y_p$ and $T$ as in the top panel are considered.
It comes out that $0.05 \lesssim n_{\mathrm{tr}} \lesssim 0.14~\mathrm{fm}^{-3}$ and there is a significant $T$- and $Y_p$-dependence.
The transition via mechanism II gives the highest values of $n_{\mathrm{tr}}$. 
In cold matter, the transition density varies between $\sat{n}/3$ and $\sat{n}/2$, depending on $Y_p$. 
A question that arises now is how close these transition densities are to those predicted by more sophisticated models, which implement in-medium and temperature modifications of nuclear cluster functionals. In Sec.~\ref{ssec:NSCrust} we shall see that for cold NS matter our results are in fair agreement with those of microscopic models.

Determination of the subsaturation density where clusterized stellar matter is superseded by homogeneous matter is not enough to build the kind of EOS tables that numerical simulations of CCSN and BNS mergers require. 
The reason is that the Coulomb interaction present in the clusterized phase and absent in the homogeneous one results in the neutron and proton chemical potentials in one phase and the other being offset, see Eq.~\eqref{eq:mugas}.
A thermodynamically stable matching between one phase and the other, both having the same $Y_p$ value, can be done by identifying the low density counterpart of the homogeneous matter just after the transition and smoothly merging them. 
The quantity we choose for identifying this low-density counterpart is the pressure.
This smooth matching consists in replacing all eNSE values in between with the results of the linear interpolation between the eNSE solution at low density and its homogeneous counterpart at high density.
A similar solution was adopted by \cite{Hempel_NPA_2010}. 
The composition of matter in this transition domain is derived by linearly diminishing the abundances of all light and heavy nuclei that exist at the low density border, without altering their mass and charge; the amount of unbound nucleons results from mass and charge conservation.

This approach reminds one of the Maxwell construction. We insist that it is only a smooth matching.
There are two arguments against interpretation as a Maxwell construction. 
First, a Maxwell construction assumes that a phase transition takes place but, as it will be demonstrated in Sec.~\ref{sec:Stability},
clusterized matter with electrons is stable.
Second, two or more phases can be put in equilibrium if all the intensive variables are equal.
This is not the case here, as only the pressure has the same value in one phase and in the other.

\section{Thermodynamic stability of stellar matter}
\label{sec:Stability}

\begin{figure}
  \centering
   \includegraphics[scale=0.43]{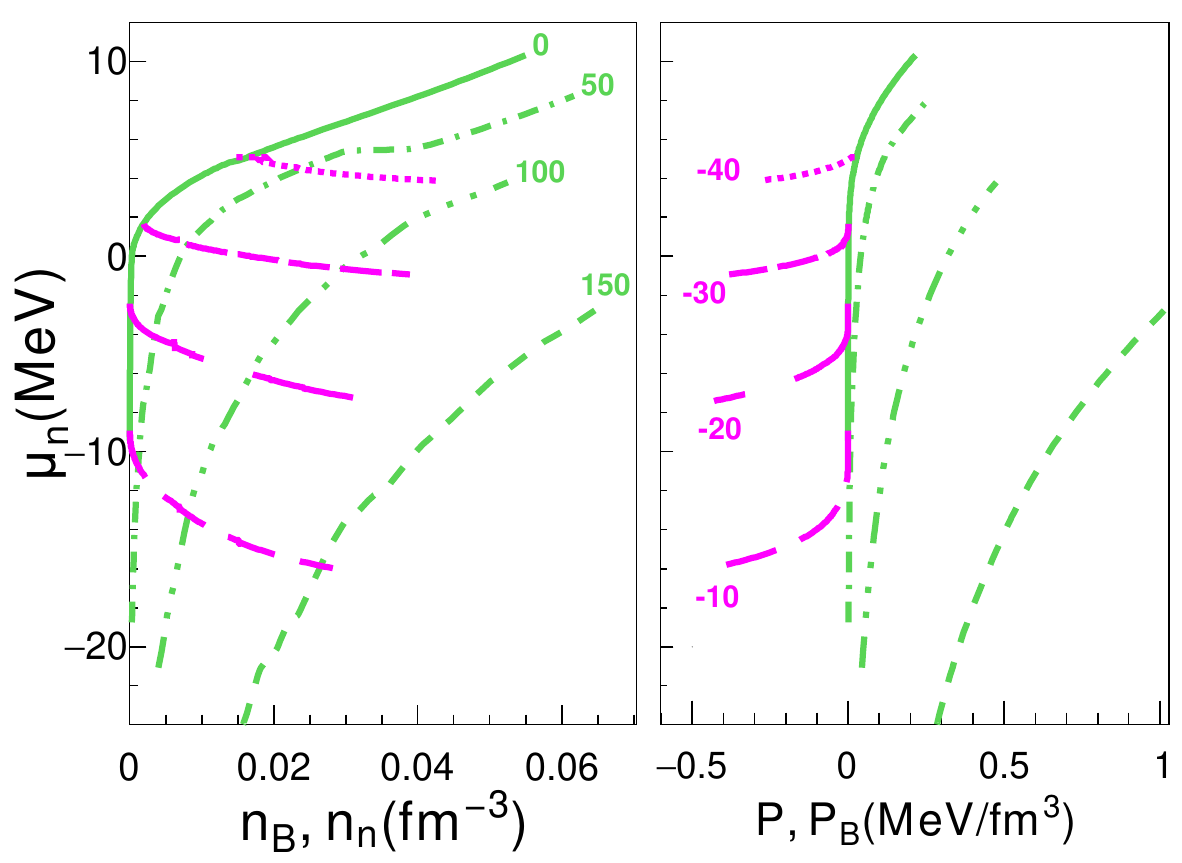}
   \caption{Thermodynamic (in)stability of cold clusterized matter.
    Green curves: properties of clusterized stellar matter with fixed lepton chemical potentials (expressed in MeV) in terms of baryon chemical potential $\mu_B(=\mu_n)$ as a function of baryonic number density $n_B$ (left panel) and total pressure $P$ (right panel).
    Magenta curves: properties of clusterized baryonic matter with fixed proton chemical potentials (expressed in MeV) in terms of neutron chemical potential $\mu_n$ as a function of neutron number density $n_n$ (left panel) and baryonic pressure $P_B$ (right panel).
    The considered interaction is BBSk1.
    \label{Fig:ThermoStab}
  }
\end{figure}

Relativistic~\citep{Mueller_PRC_1995,Liu_PRC_2002,Providencia_PRC_2006,Avancini_PRC_2006} and non-relativistic~\citep{Ducoin_NPA_2006,Ducoin_NPA_2007,Rios_NPA_2010} mean-field calculations as well as {\em ab initio} approaches~\citep{Vidana_PLB_2008,Rios_PRC_2008,Carbone_PRC_2018} demonstrate that over a large density domain dilute homogeneous NM without Coulomb interactions manifests density-driven instabilities.
According to \cite{Typel_PRC_2010,Raduta_PRC_2009,Raduta_PRC_2010}, the same is the case of dilute clusterized NM.

The thermodynamic stability of clusterized NM is addressed in Fig.~\ref{Fig:ThermoStab} (left panel) in terms of neutron chemical potential $\mu_n$ as a function of neutron density $n_n$ for constant values of proton chemical potential $\mu_p$~\citep{Ducoin_NPA_2006}, magenta curves.
For simplicity, only the case of $T=0.1~\mathrm{MeV}$ is addressed.
For all considered values of $\mu_p$ and over the entire illustrated density domain, $\mu_n(n_n)$ decreases, which indicates that the system is thermodynamically unstable. 
The set of green curves in Fig.~\ref{Fig:ThermoStab} corresponds to clusterized stellar matter, that is, a mixture of nuclei, nucleons, electrons, and photons with a positive charge of protons fully compensated by the negative charge of electrons. In this case, the baryon chemical potential $\mu_B(=\mu_n)$ is plotted as a function of the baryon density $n_B$ for constant values of the lepton chemical potential $\mu_L$~\citep{Ducoin_NPA_2007}. 
The curve with $\mu_L=0$ corresponds to beta-equilibrated matter.
None of the $\left.\mu_B (n_B)\right|_{\mu_L}$-curves decreases with density, which means that no instability occurs in that domain of density and isospin asymmetry. 
This demonstrates that the necessity for Maxwell constructions does not arise at $T=0.1~\mathrm{MeV}$.
For homogeneous NM, the phase instability domain shrinks with temperature; the same holds true for clusterized NM without Coulomb~\citep{Raduta_PRC_2010}. 
Considering that cold clusterized stellar matter does not feature any instability, it is reasonable to assume that the matter is stable at higher temperatures, too.  

Additional information is presented on the right panel, which shows the curves of nuclear pressure $P_B$ vs $\mu_n$ for constant values of $\mu_p$ (magenta) and the curves of total pressure $P$ vs $\mu_B$ for constant values of $\mu_L$ (green). $P_B$ is obtained from the total pressure $P$ stored in \textsc{CompOSE} tables after subtracting the electron and photon contributions.
The magenta (green) family of curves resembles the corresponding curves of homogeneous nuclear (stellar) matter, i.e., are convex (concave) whenever matter is unstable (stable) with respect to phase separation.

Notice that the conclusions of this section are not incompatible with sub-saturated NM with electrons being unstable with respect to finite size density fluctuations~\citep{Baym_NPA_1971,Pethick_NPA_1995,Ducoin_NPA_2007}. On the contrary, present results demonstrate that after clusterization due to periodic microscopic instabilities the matter becomes stable.

\section{Results}
\label{sec:Results}

This section is devoted to the results of our eNSE model. 
First, we investigate the composition of the NS crusts and its dependence on the effective interaction.
A systematic comparison with the predictions of the extended Thomas-Fermi plus Strutinsky integral (ETFSI) method by \cite{Pearson_MNRAS_2018} is also carried out.
Then, we turn to the finite-temperature regime.
The composition and energetics of sub-saturated matter are investigated for a wide range of temperatures and proton fractions.

\subsection{Neutron star crusts}
\label{ssec:NSCrust}

\begin{figure}
  \centering
  \includegraphics[scale=0.38]{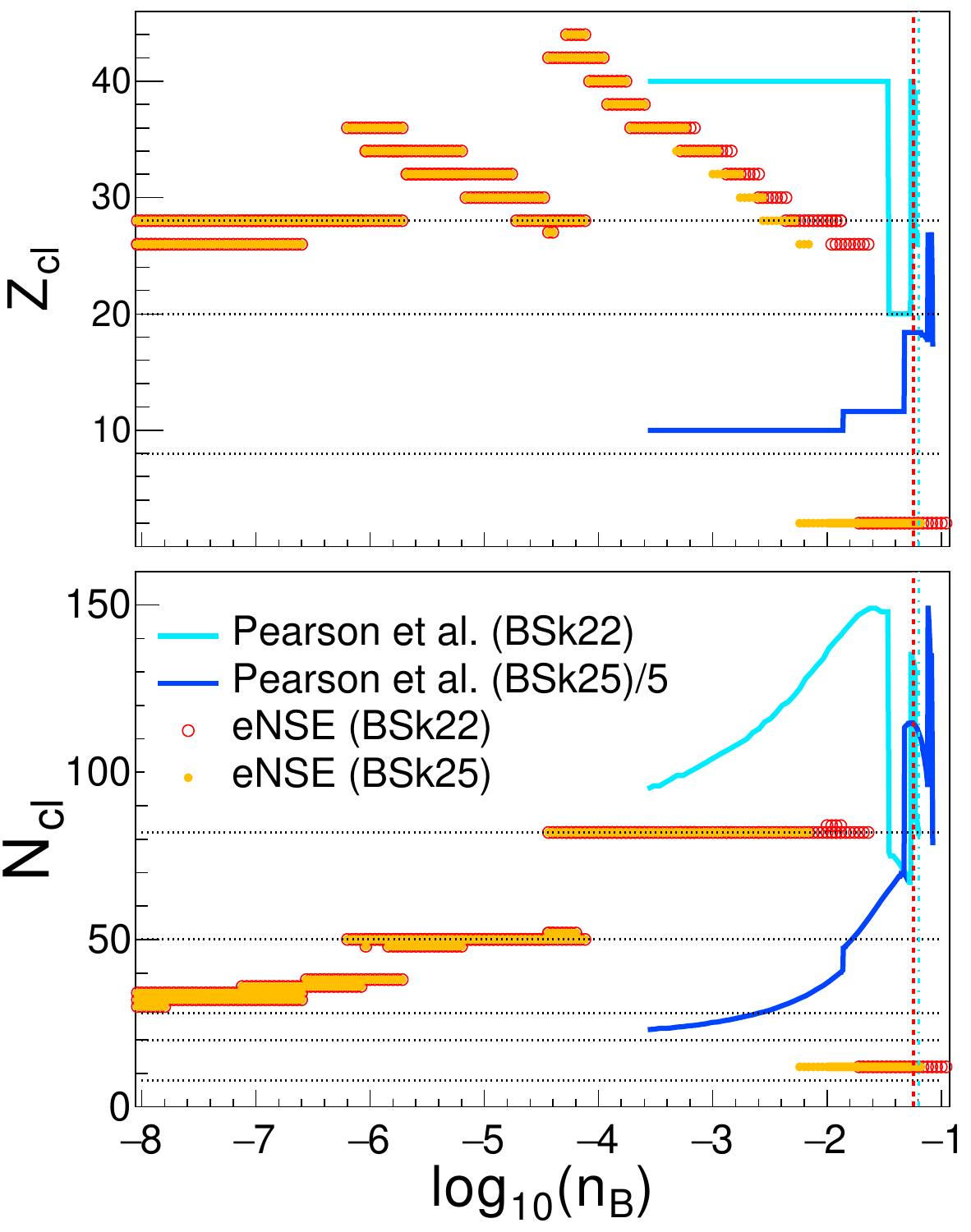}
  \caption{Composition of an NS crust as a function of baryon number density in terms of
    numbers of neutrons (bottom panel) and protons (top panel) of nuclei.    
    eNSE results obtained in this work for $\beta$-equilibrated matter with $T=0.1~\mathrm{MeV}$ are compared with results of \cite{Pearson_MNRAS_2018}. 
    The considered effective interactions are BSk22 and BSk25~\citep{BSk22-BSk26}.
    Notice that BSk25 results of \cite{Pearson_MNRAS_2018} are scaled by a factor of $1/5$.
    Vertical dashed lines mark the transition to homogeneous matter for BSk22:
    $n_{\mathrm{tr}}=5.70 \cdot 10^{-2}~\mathrm{fm}^{-3}$ and $6.35 \cdot 10^{-2}~\mathrm{fm}^{-3}$
    for the eNSE results and the results of \cite{Pearson_MNRAS_2018}, respectively.
    Horizontal dotted lines correspond to magic numbers, Z=8, 20, 28 and N=8, 20, 28, 50, 82.
  }
  \label{Fig:NS_compo}
\end{figure}

\begin{figure}
  \centering
  \includegraphics[scale=0.38]{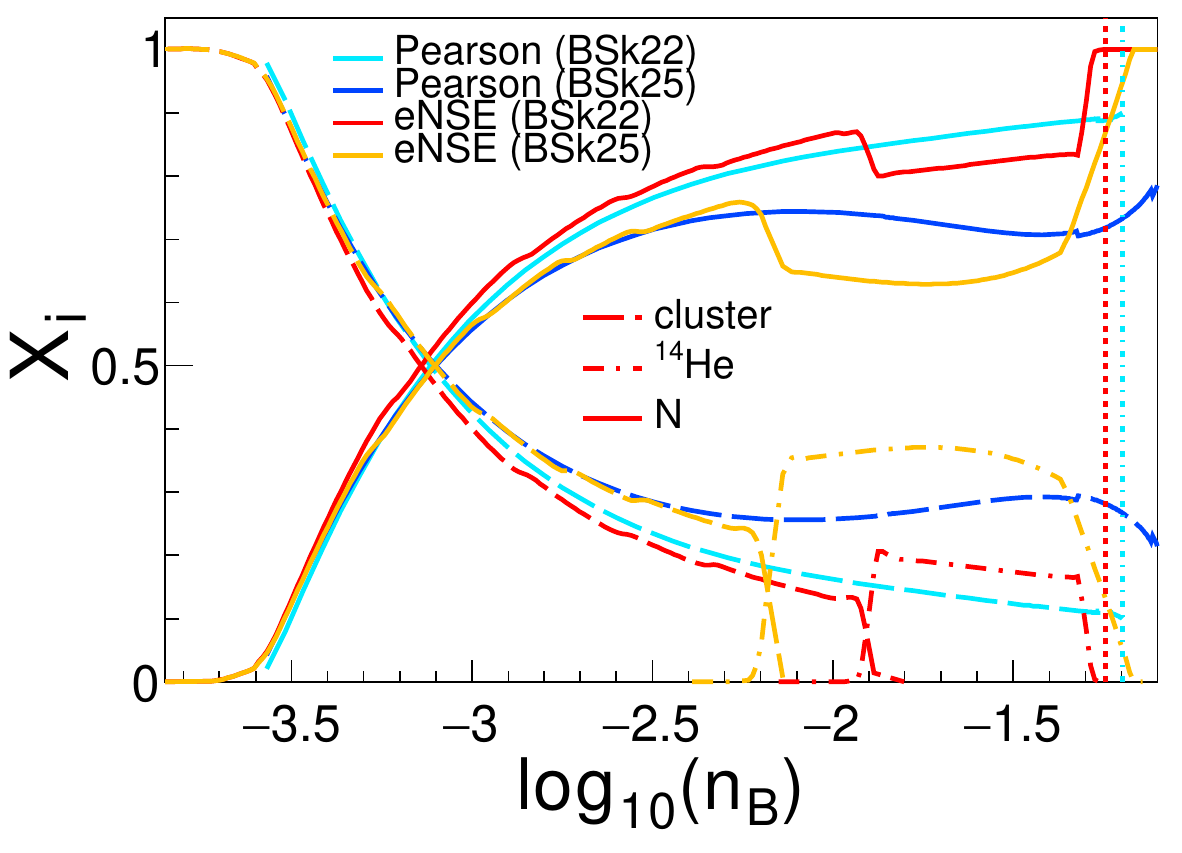}
  \includegraphics[scale=0.38]{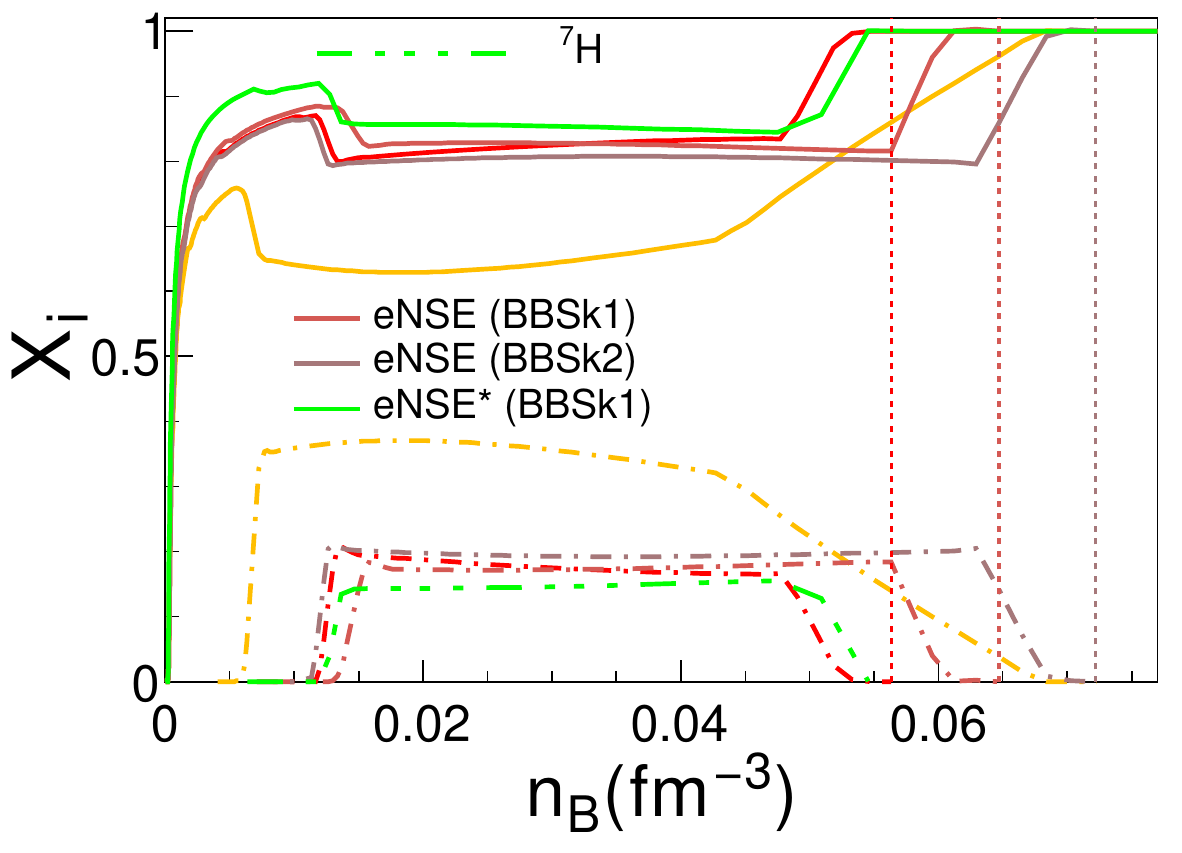}  
  \caption{Mass fractions of various species present in the inner crust and their dependence on the effective interaction as functions of baryon number density. eNSE results are for $T=0.1~\mathrm{MeV}$.
Top panel: eNSE predictions are confronted with the zero-temperature results of \cite{Pearson_MNRAS_2018}.
Bottom panel: mass fractions of $^{14}$He, $^7$H and unbound nucleons, as predicted by eNSE.
The considered effective interactions are BSk22 and BSk25 from \citep{BSk22-BSk26} and BBSk1 and BBSk2 proposed here. As in Fig.~\ref{Fig:NS_compo}, vertical dashed lines mark the transition to homogeneous matter.
  }
  \label{Fig:NS_massfrac}
\end{figure}

\begin{figure}
  \centering
  \includegraphics[scale=0.4]{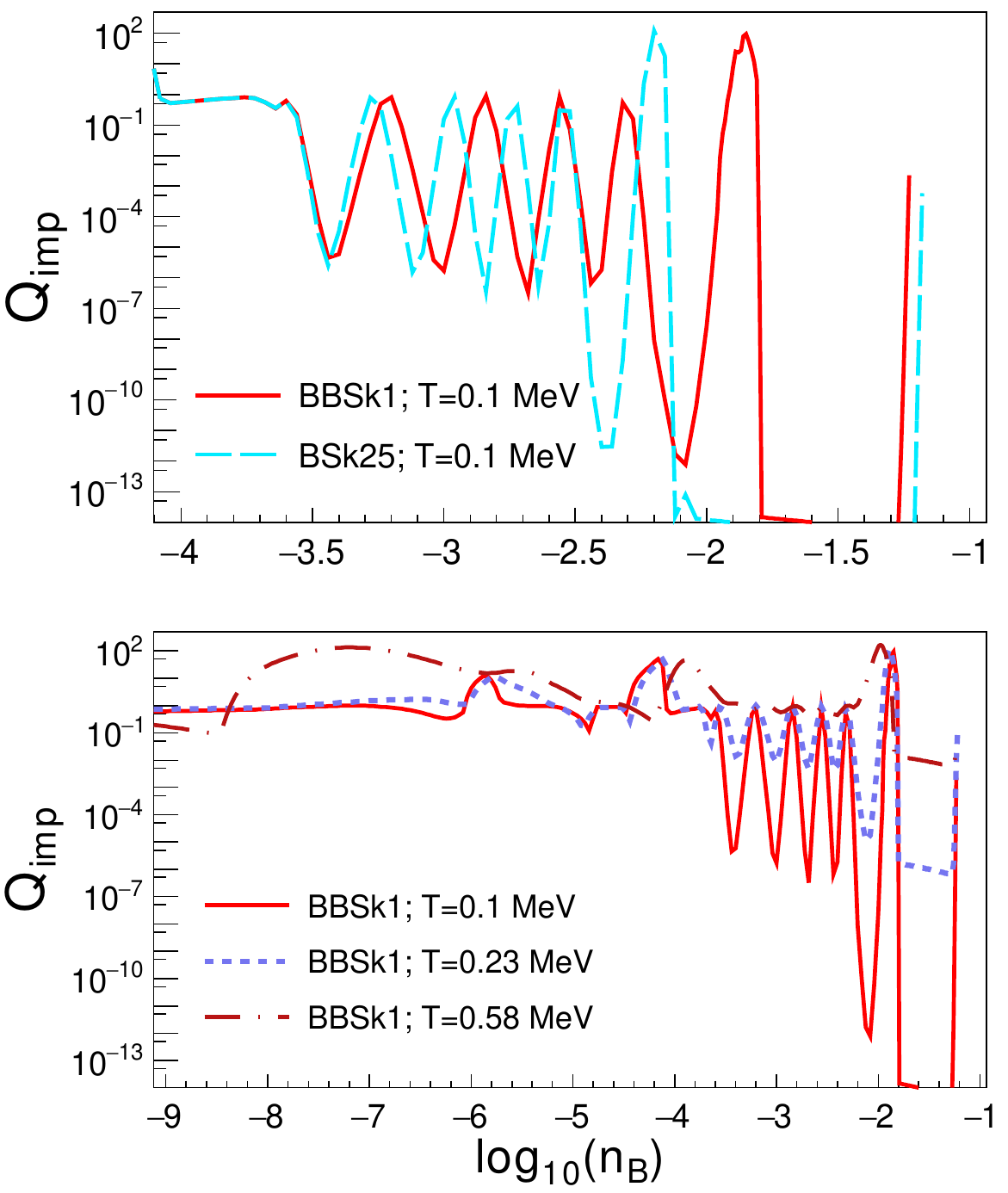}
  \caption{Impurity parameter $Q_{\mathrm{imp}}$ in the NS crust.
    Bottom panel: predictions of BBSk1 at various temperatures.
    Top panel: effective interaction effects at $T=0.1~\mathrm{MeV}$.
  }
  \label{Fig:Qimp}
\end{figure}

Here, we investigate the predictions of our eNSE model in the limit of low temperatures.
Because $T=0.1~\mathrm{MeV}$ considered in this subsection falls outside the validity domain of the nuclear statistical equilibrium, some explanations are necessary.

The reason why we expect NSE to give a fair description of the composition is that, as the temperature approaches zero, NSE distributions are practically reduced to a single nucleus~\citep{Hempel_NPA_2010,Gulminelli_PRC_2015}. 
Then, an NSE state is by construction a state of equilibrium. 
In the limit of vanishing temperatures, that state corresponds to the minimum energy of the system. 
If the controlled proton fraction is tuned to comply with $\mu_n=\mu_p+\mu_e$, NSE should provide a good approximation for the NS crust. 

The composition of cold $\beta$-equilibrated matter is addressed in Fig.~\ref{Fig:NS_compo} in terms of neutron and proton numbers of the nuclear species present at various depths inside the NS crust. The effective interactions considered are BSk22 and BSk25~\citep{BSk22-BSk26}.
It turns out that the crust composition is stable over density domains of various widths and when it gets modified the changes are drastic.
This is the well-known outcome of shell and odd-even effects in nuclear masses.
Indeed, one can recognize the $Z=28$ and $N=50,~82$ magic numbers along with many even-$Z$ numbers in the inner crust.
The constraints of the AME2020~\citep{AME2020} and DZ10~\citep{Duflo_PRC_1995} nuclear mass tables, which we employ and which limit the crust composition, are such that the size of the cluster does not always increase with density.
In particular, over $5 \cdot 10^{-7} ~\mathrm{fm}^{-3}\lesssim n_B \lesssim 8 \cdot 10^{-5}~\mathrm{fm}^{-3}$ and
$4 \cdot 10^{-5} ~\mathrm{fm}^{-3}\lesssim n_B \lesssim 10^{-2}~\mathrm{fm}^{-3}$
nuclei in the crust have $N=50$ and $N=82$ while their charge number decreases from 36 to 28 and from 42 to 28, respectively.
Over $6.3 \cdot 10^{-7} ~\mathrm{fm}^{-3}\lesssim n_B \lesssim 2.5 \cdot 10^{-6}~\mathrm{fm}^{-3}$ and
$4 \cdot 10^{-5} ~\mathrm{fm}^{-3}\lesssim n_B \lesssim 8 \cdot 10^{-5}~\mathrm{fm}^{-3}$ two species with significantly different $Z$- and $N$-numbers are in competition,and, for some densities, neighboring nuclides are present, too.
The novel feature predicted by our model is the occurrence of a thick $^{14}$He shell at the bottom of the crust. This nuclide is present in the DZ10 table~\citep{Duflo_PRC_1995}; its one- and two-neutron separation energies are $-1.29~\mathrm{MeV}$ and $-5.47~\mathrm{MeV}$, respectively.
The occurrence of light neutron-rich nuclides in the innermost layers of the crust was checked by limiting the pool of NSE nuclei to those present in the AME2020~\citep{AME2020} table only. The results indicate that, in this case, a layer of $^7$H is present instead of $^{14}$He, see the discussion below.

Even if non-spherical degrees of freedom are not explicitly implemented in our work, we can speculate that the onset of light neutron-rich nuclides might challenge the pasta phases, should the model account for them.
A detailed study of the consequences of such layers on the mechanical and transport properties of the crust is beyond the scope of this article.
For the issue of mechanical stability, see \cite{Beznogov_PRL_2025}; some comments on the impurity parameter are given below.
Effective interaction effects manifest in the inner crust, but the representation in Fig.~\ref{Fig:NS_compo} does not allow one to quantify their amplitude. It is nevertheless clear that much stronger effective interaction effects would manifest should larger and more neutron-rich nuclei be allowed to exist and their binding energy to reflect current uncertainties in the isovector channel. 

For the sake of comparison, the results of ETFSI calculations with the BSk22 and BSk25 interactions~\citep{BSk22-BSk26} by \cite{Pearson_MNRAS_2018} are plotted, too. For BSk22 (BSk25) they show that up to $n_B=3.4 \cdot 10^{-2}~\mathrm{fm}^{-3}$ ($n_B=1.4 \cdot 10^{-2}~\mathrm{fm}^{-3}$) only Zr (Sn) is present in the inner crust and its neutron enrichment increases steeply and smoothly with density; for BSk22, a Ca-layer exists over
$3.5 \cdot 10^{-2}~\mathrm{fm}^{-3} \lesssim n_B \lesssim 5.3 \cdot 10^{-2}~\mathrm{fm}^{-3}$; shell effects manifest only in the charge number.
The size of the nuclear cluster predicted by ETFSI is much larger than the one we obtain.
 
The vertical dashed lines mark the crust-core transition densities for BSk22. In spite of performing calculations on a grid and limiting the crust composition to species present in the mass tables, the value we obtain for the crust-core transition density is remarkably close to the value obtained by \cite{Pearson_MNRAS_2018}.

More insight into the inner crust composition is provided in Fig.~\ref{Fig:NS_massfrac}. 
The top panel depicts the mass fractions of nuclei and unbound nucleons as functions of number density. The contributions of $^{14}$He are indicated with dot-dashed curves.
Despite the stark differences with the predictions of \cite{Pearson_MNRAS_2018} in regard to the properties of nuclei, the mass sharing between the nuclei and nucleons is fairly similar in our model and the ETFSI model of \cite{Pearson_MNRAS_2018}, at least until the emergence of  $^{14}$He. This suggests that the gross behavior in the inner crust is governed by the effective interaction.
The nucleation of $^{14}$He in the deep crust entails a steep depletion of the gas of unbound nucleons with possible effects on $^1S_0$ superfluidity.
The effective interaction dependence of the inner crust composition is further investigated in the bottom panel, where, in addition to the results corresponding to BSk22 and BSk25, we also include the results corresponding to BBSk1 and BBSk2 forces.
Results corresponding to the case where the NSE pool of nuclei is limited to AME2020 table only (eNSE$^*$) are illustrated, too. One can see that $^{7}$H has the same onset density and slightly smaller mass fraction ($\approx 14\%$ vs $\approx 20\%$) compared to $^{14}$He.
The onset density of $^{14}$He and the relative mass sharing between $^{14}$He and the nucleon gas depend on the underlying interaction.
The same holds for the width of the $^{14}$He layer and the transition density to homogeneous matter.

The crust of an isolated NS forms during the neo-NS evolutionary phase~\citep{Beznogov_2020}. This stage begins at the end of the proto-NS epoch, when the star’s temperature drops below $\approx 2~\mathrm{MeV}$ and its material becomes transparent to neutrinos, and lasts for about a day.
Two episodes are of particular interest for the matters discussed here.
One corresponds to the instance when NSE freezes out and marks the beginning of the epoch when only a limited subset of nuclear reactions can change the composition of matter.
The other corresponds to the crystallization of the crust.
The freeze-out temperature of NSE depends on the density and, even if further composition changes are possible, its value plays an important role for crust composition and thermal properties.
The crystallization temperature of a Coulomb plasma depends on the ion's charge squared~\citep{Chamel_LivRev_2008}. This means that the inner crust of BSk22 by \cite{Pearson_MNRAS_2018} will have a crystallization temperature higher by a factor of one hundred compared to our inner crust with $^{14}$He layer.
In the absence of dynamical simulations of crust formation, it is difficult to say which of these instances occurs first and what exactly the crust composition and transport properties are. As such, only statistical estimates can be made. A useful quantity to this aim is the impurity parameter~\citep{Potekhin_AA_2021},
\be
Q_{\mathrm{imp}}=\sum_j y_j \left(Z_j -\left< Z\right> \right)^2,
\ee
where $\left< Z\right>=\sum_j y_j Z_j$ is the mean charge number, $y_j=Y_j/\sum_i Y_i$ and $Y_i$ stands for the yield per unit volume of species $i$.

Figure~\ref{Fig:Qimp} (bottom panel) illustrates the density dependence of $Q_{\mathrm{imp}}$ for $0.1~\mathrm{MeV} \leq T \leq 0.58~\mathrm{MeV}$. For the lowest temperature, $Q_{\mathrm{imp}} \approx 0$ only in the highly pure $^{14}$He layer.
Over $10^{-9}~\mathrm{fm}^{-3} \lesssim n_B \lesssim 10^{-6}~\mathrm{fm}^{-3}$, $Q_{\mathrm{imp}}$ changes
smoothly between $\approx 0.3$ and $\approx 3$; these values reflect the non-pure but rather stable composition, see Fig.~\ref{Fig:NS_compo}.  
Over $10^{-4}~\mathrm{fm}^{-3} \lesssim n_B \lesssim 10^{-2}~\mathrm{fm}^{-3}$, $Q_{\mathrm{imp}}$ fluctuates widely over fourteen orders of magnitude;
this behavior reflects sudden and drastic changes in composition.
With the increase of temperature, more species are present and the composition changes less drastically with density. 
Consequently, the $Q_{\mathrm{imp}}$ oscillations at high densities are dumped, while some oscillations appear at low densities; the lowest values correspond to the $^{14}$He layer.
The upper panel addresses the effective interaction dependence of this quantity in the inner crust. 
It comes out that while the amplitude of the oscillations seems to be preserved, some offset (``phase shift'') exists.

While the oscillations of $Q_{\mathrm{imp}}$ are well known in the literature (see, e.g., \cite{Carreau_2020b,Fantina_AA_2020,Potekhin_AA_2021}), our results are somewhat different.  
We have smoother behavior in the outer crust and strong oscillations in the outer layers of the inner crust.
However,  in the deeper layers of the inner crust, after the onset of  $^{14}$He, $Q_{\mathrm{imp}}$ becomes almost density independent, which is not the case of other models~\citep{Carreau_2020b,Potekhin_AA_2021}. 
Moreover, $Q_{\mathrm{imp}}$ becomes independent of the chosen effective interaction, which is in strong contrast with the results of \cite{Potekhin_AA_2021}.
Thus, if the existence of the almost pure $^{14}$He or $^7$H layer is proven, it will be a step forward towards a universal (i.e., EOS-independent)  description of the transport properties of the deep layers of the  inner crusts. 

\subsection{Finite temperature results}
\label{ssec:FiniteT}

This subsection presents an overview of the composition and energetics of warm and dilute NM.
A series of quantities corresponding to a fixed value of temperature (proton fraction) and different values of the proton fraction (temperature) will be plotted as functions of density.
The particular values of these quantities are chosen to highlight the most interesting features.
Also, they correspond to mesh points, which keeps the results free from inherent interpolation-related effects.

\subsubsection{Composition}
\label{sssec:Compo}

\begin{figure*}
  \centering
  \includegraphics[scale=0.45]{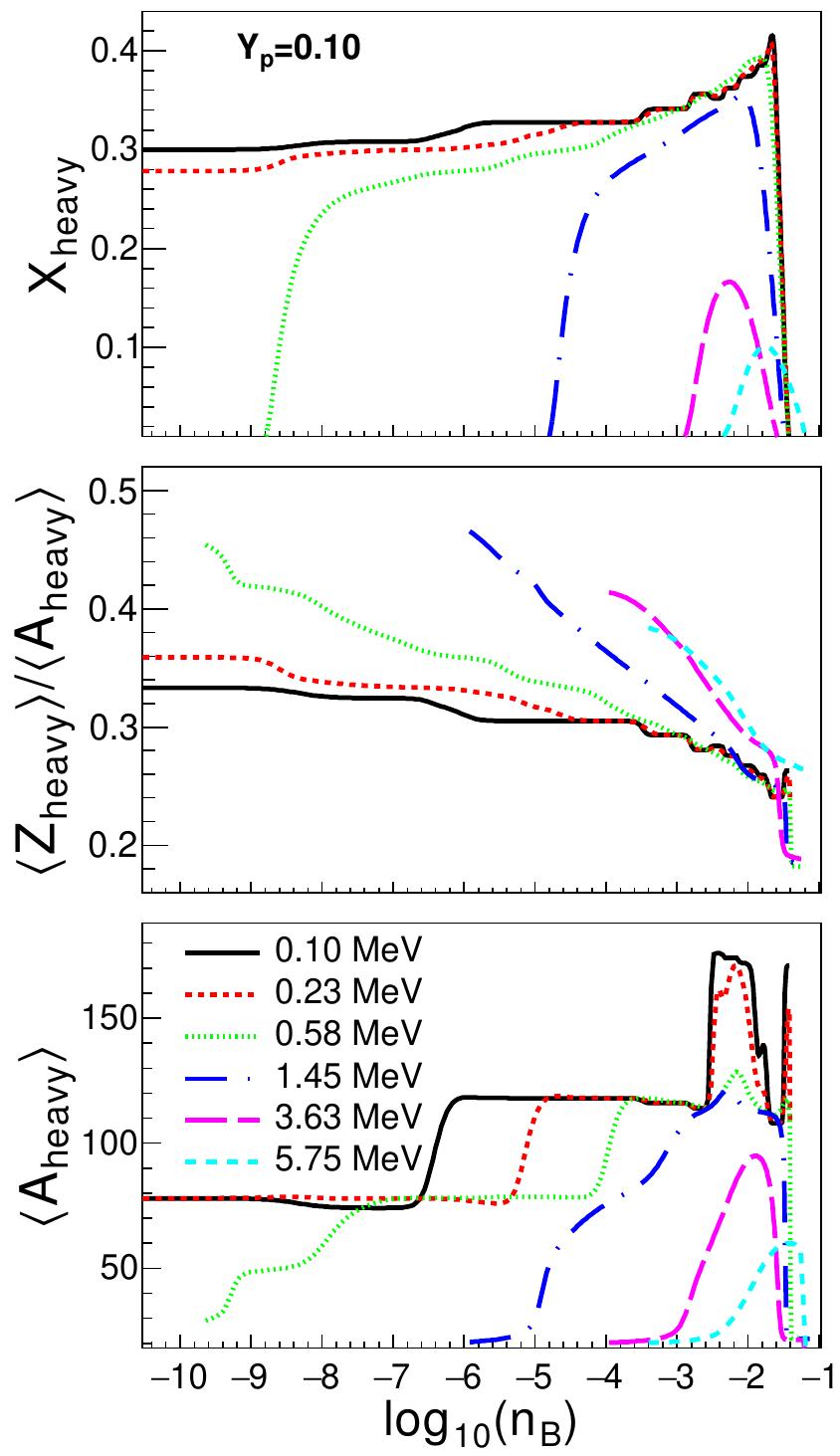}
  \includegraphics[scale=0.45]{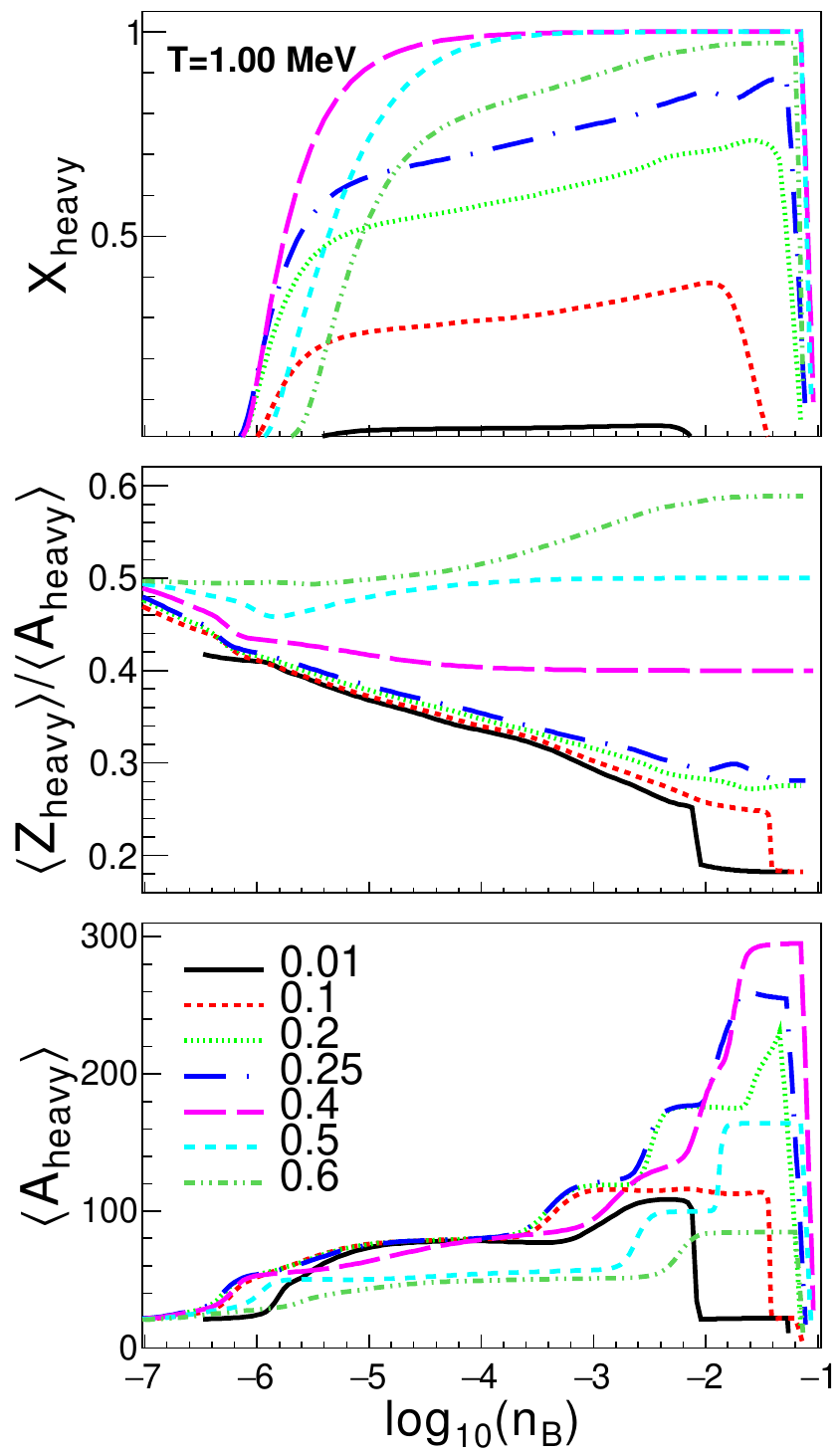}
  \caption{Average mass of nuclei with $A \geq 20$, the corresponding $\langle Z \rangle/\langle A \rangle$ as well as the mass fraction they bind as functions of baryonic particle density for $Y_p=0.1$ and different temperatures (left panels) and $T=1~\mathrm{MeV}$ and different values of $Y_p$ (right panels). For more details, see the legend.
    The considered effective interaction is BBSk1.}
  \label{Fig:HeavyCl}
\end{figure*}

\begin{figure*}
  \centering
  \includegraphics[scale=0.45]{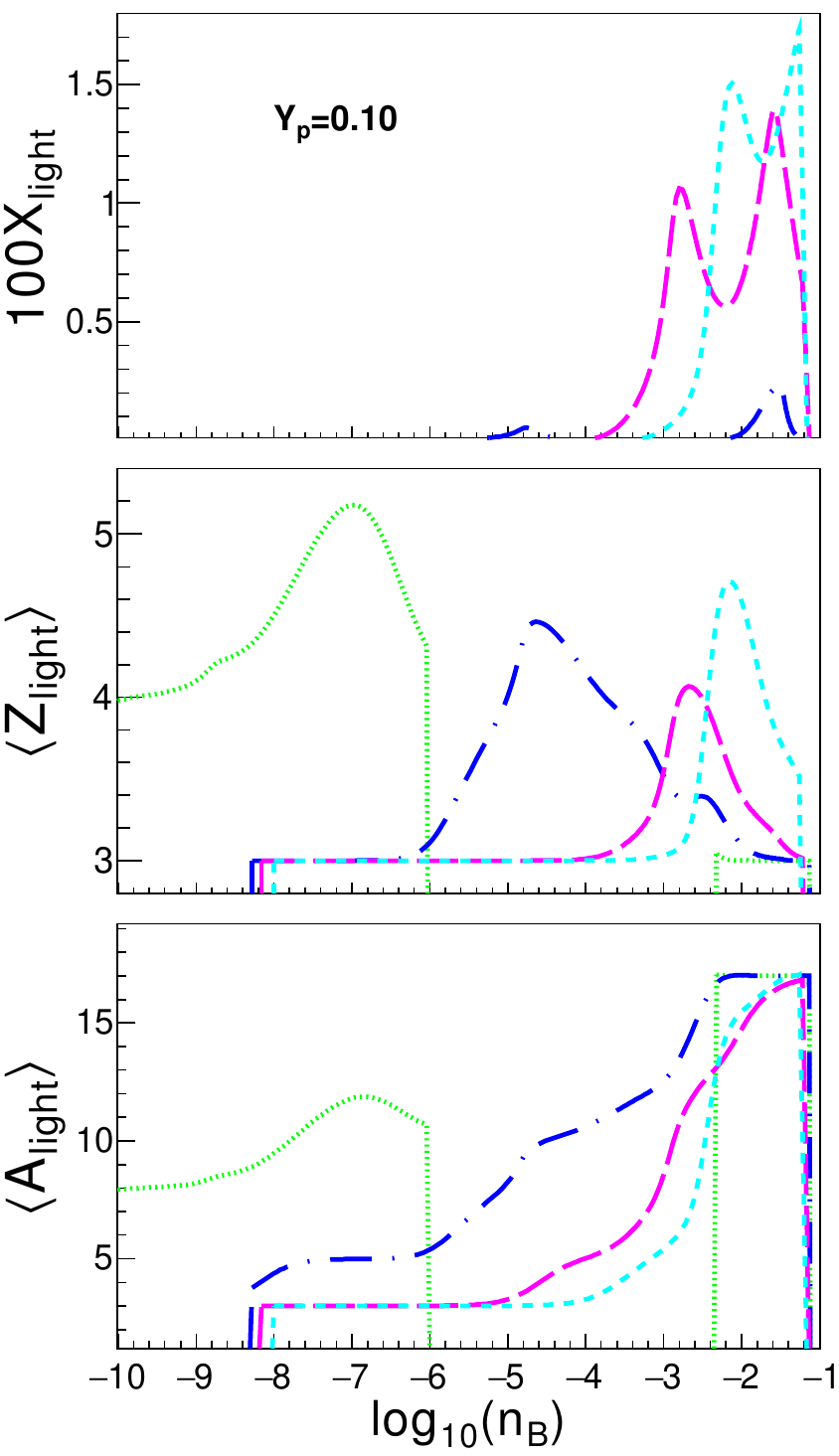}
  \includegraphics[scale=0.45]{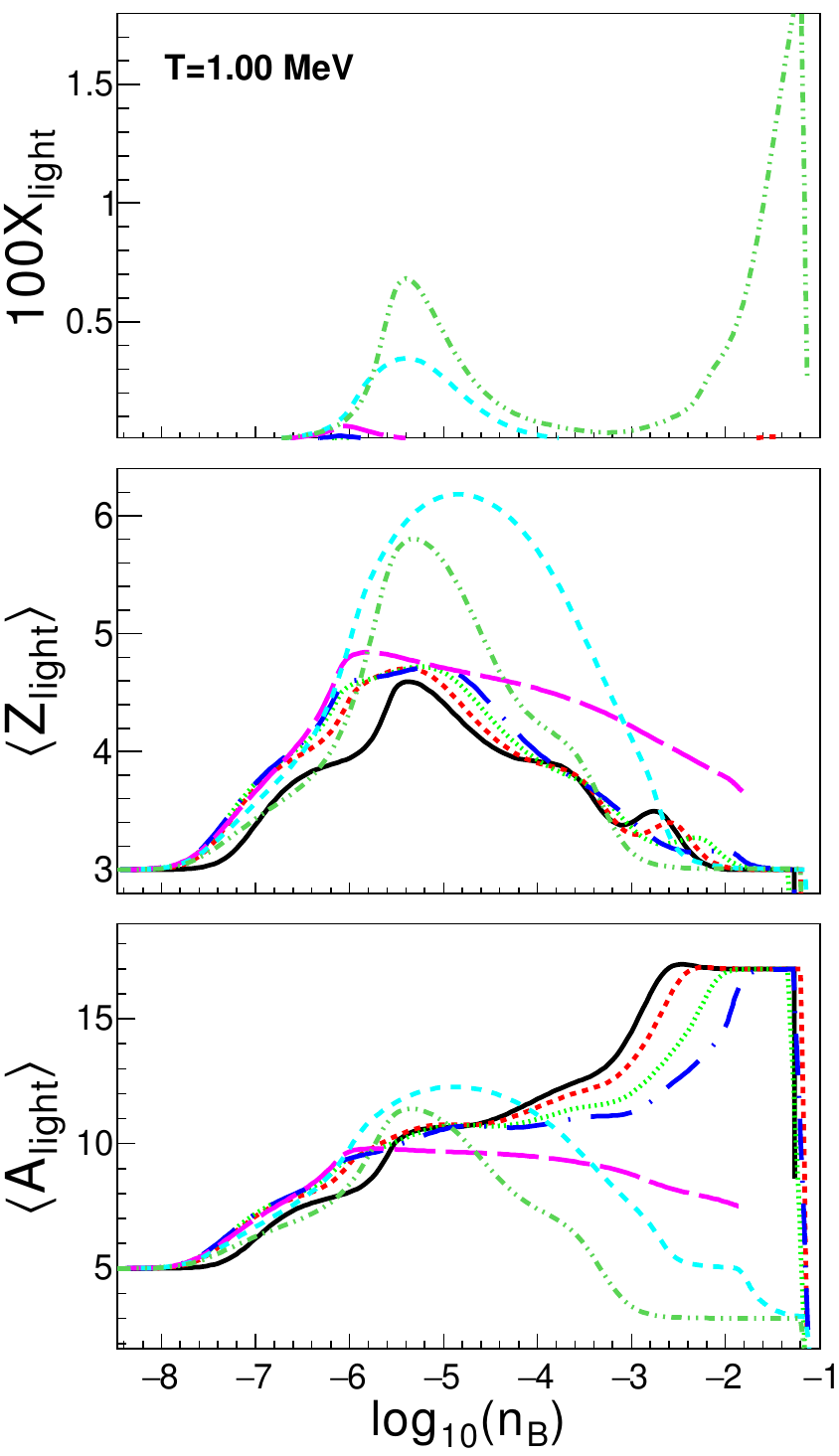}
  \caption{The same as in Fig.~\ref{Fig:HeavyCl} but for nuclei with $Z \geq 3$ and $A <20$. Notice that the $X$-range is not the same as in  Fig.~\ref{Fig:HeavyCl}.}
  \label{Fig:LightCl}
\end{figure*}

\begin{figure*}
  \centering
  \includegraphics[scale=0.33]{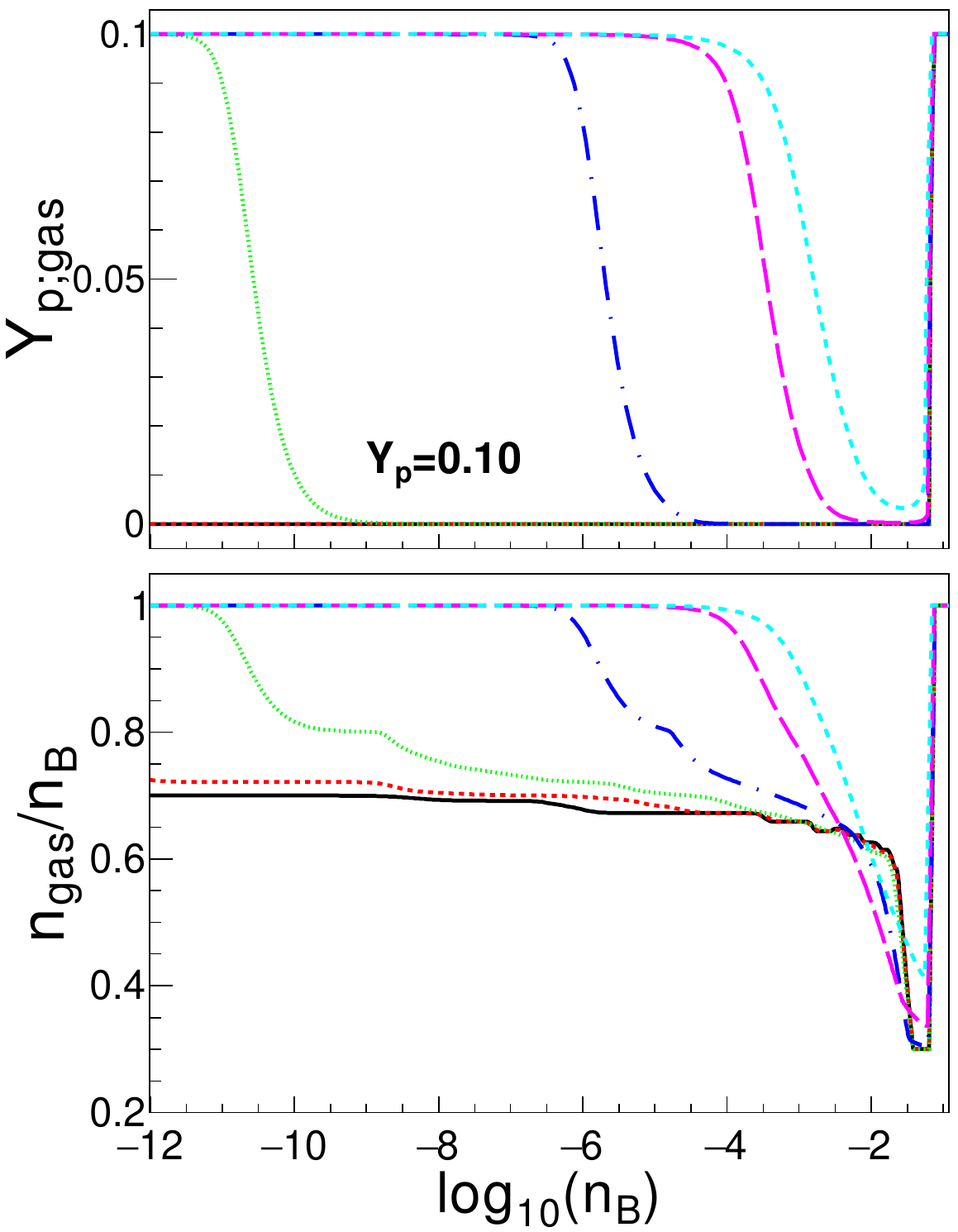}
  \includegraphics[scale=0.33]{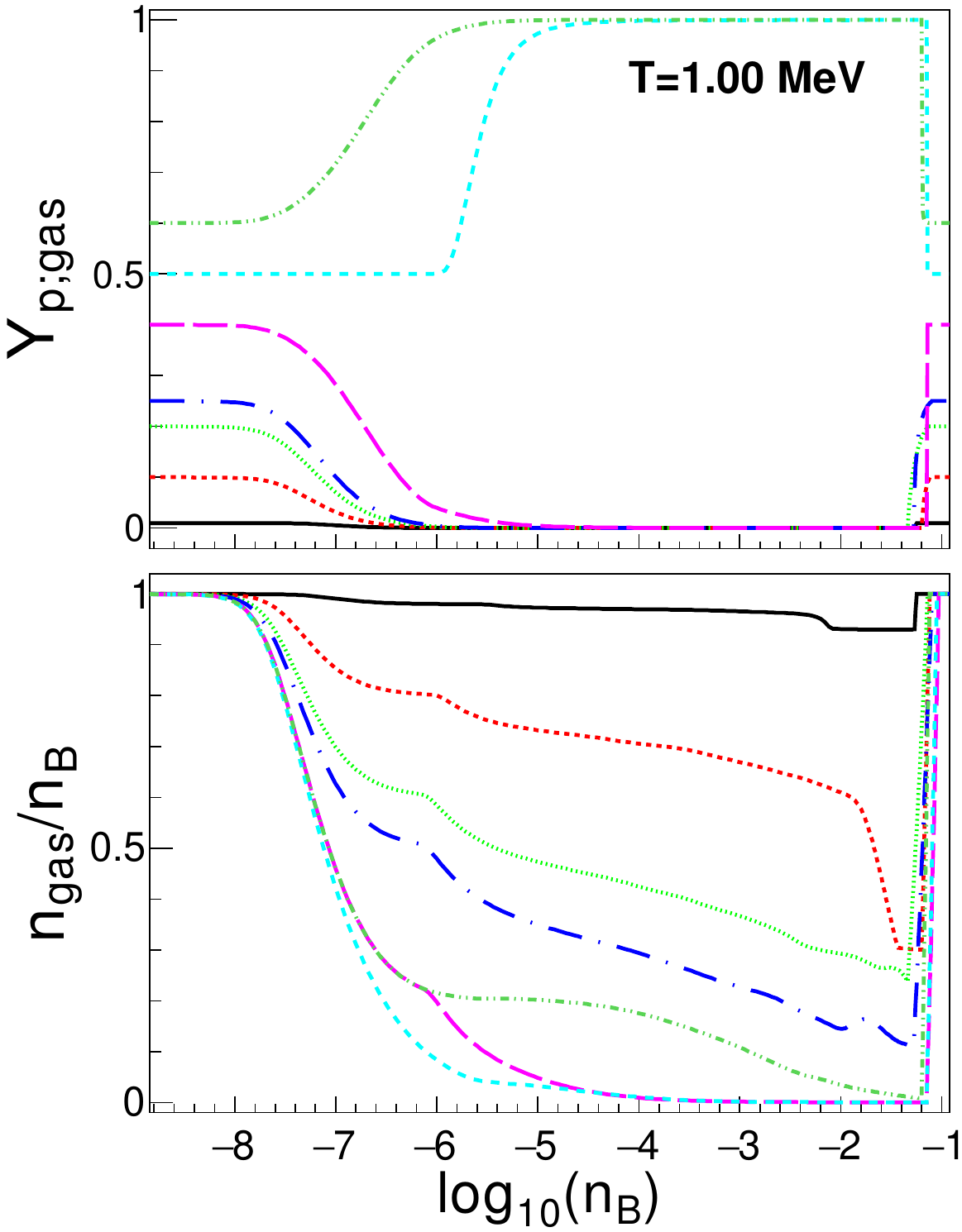}
  \caption{Mass fraction of unbound nucleons (bottom panels) and their proton fraction (top panels) as functions of baryonic particle density for the cases considered in Fig.~\ref{Fig:HeavyCl}. For the legend, see Fig.~\ref{Fig:HeavyCl}. Notice that the $X$-range is not the same as in  Fig.~\ref{Fig:HeavyCl}.}
  \label{Fig:Gas}
\end{figure*}

\begin{figure}
  \centering
  \includegraphics[scale=0.35]{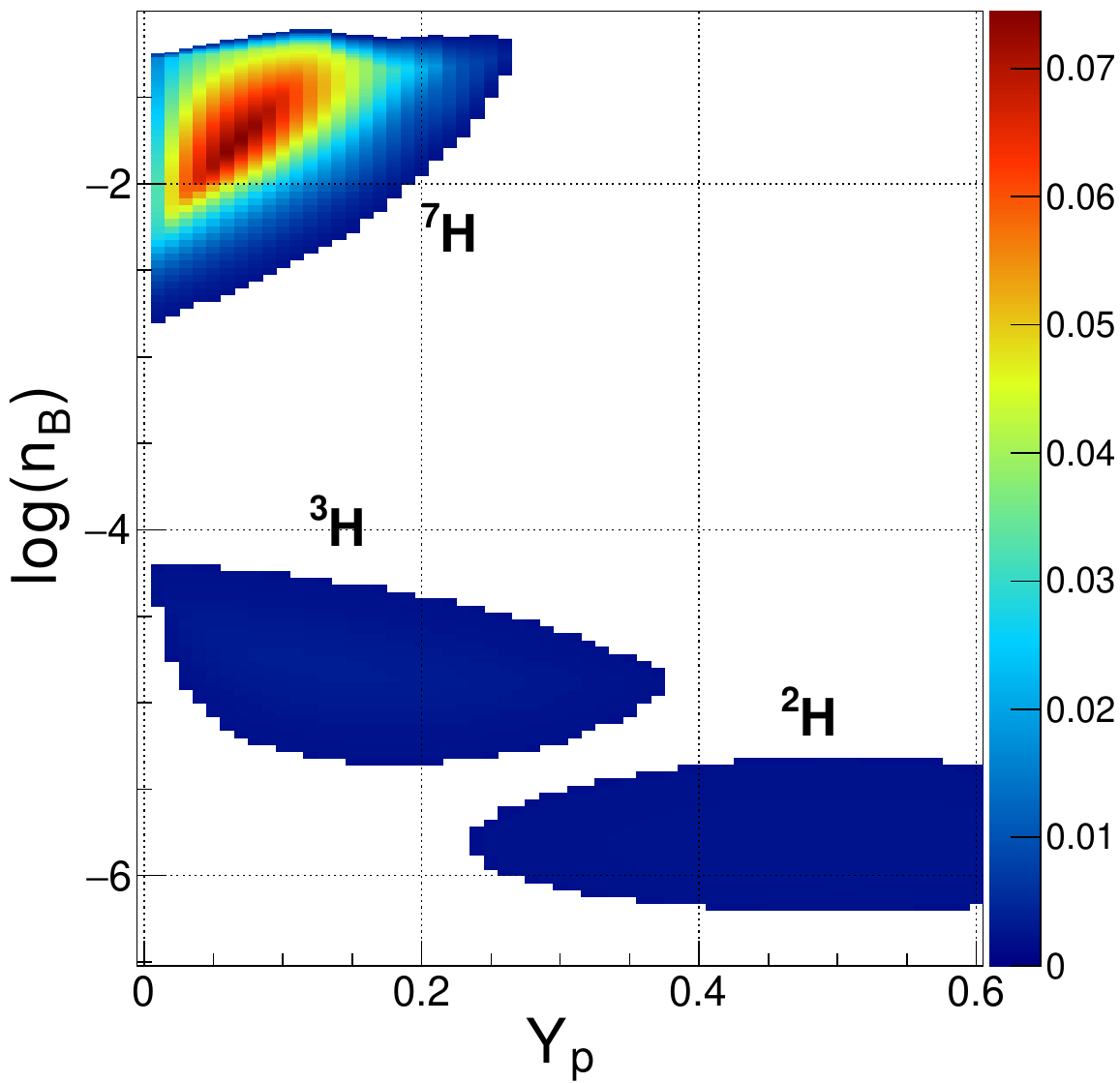}
  \includegraphics[scale=0.35]{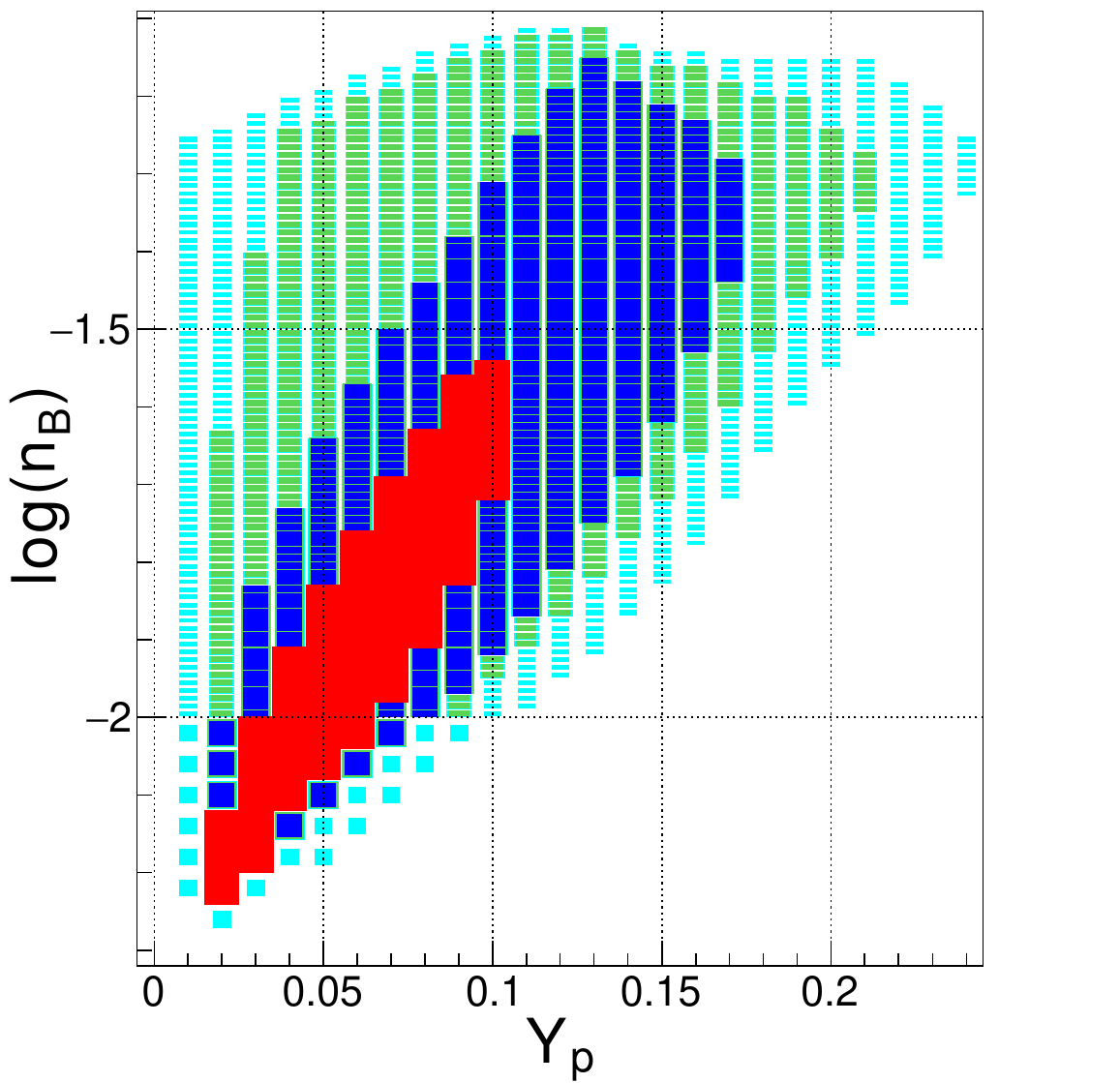}
  \caption{Domains of $n_B-Y_p$ where at $T=1.44~\mathrm{MeV}$ the mass fractions of H isotopes (top panel) and
    $^{11}$He (red), $^{12}$He (blue), $^{13}$He (green), $^{14}$He (cyan) (bottom panel) exceed 0.2\% and 1\%, respectively.
    The color scale in the top panel corresponds to the mass fraction bound in each of the considered H isotopes.
    The effective interaction is BBSk1.}
  \label{Fig:HandHe}
\end{figure}

In NSE, the clusterized component of NM accommodates a wealth of nuclei, whose number virtually equals 
the number of nuclei included in the mass tables.
Some of these nuclei have negligible yields, while others are abundant. 
The number of nuclei that appear in significant amounts depends on the thermodynamic conditions and intrinsic nuclear properties. Both aspects were extensively discussed by \cite{Hempel_NPA_2010,Raduta_PRC_2010,Gulminelli_PRC_2015,Raduta_NPA_2019,Blinnikov_AA_2011,Furusawa_ApJ_2011,Furusawa_ApJ_2013,Furusawa_JPG_2017,Furusawa_NPA_2017} and will not be repeated here. 
The only aspect we wish to remind is that whenever the nuclear distributions are multi-modal, which is typically the case at relatively low temperatures and proton fractions exceeding a certain threshold, the average numbers of neutrons and protons of nuclei in a category might not be representative of any of the nuclei that appear in large amount.

With this caveat, the composition of NM is discussed here in terms of average mass and proton numbers, 
\ba
\langle A_{\cal{C}} \rangle=\frac{\sum_{i \in \cal{C}} A_i Y_i}{\sum_{i \in \cal{C}} Y_i}, ~
\langle Z_{\cal{C}} \rangle=\frac{\sum_{i \in \cal{C}} Z_i Y_i}{\sum_{i \in \cal{C}} Y_i},
\ea
and mass fraction,
\be
X_{\cal{C}}=\frac{\sum_{i \in \cal{C}} A_i Y_i}{n_B},
\ee
of two generic species for which relative mass abundance is reported in the tables contributed to \textsc{CompOSE},
i.e., ``heavy" nuclei (${\cal C}=\{A,Z \}$ with $A \geq 20$)
and ``light" nuclei (${\cal C}=\{A,Z \}$ with $Z \geq 3$ and $A <20$).
As is Sec.~\ref{ssec:NSCrust}, in the equations above, $Y_i$ stands for the yield per unit volume of species $i$.
The properties of the gas of unbound nucleons as well as the abundances of nuclides with $Z=1$ and 2 and $A \geq 2$, also reported in the tables contributed to \textsc{CompOSE}, are addressed, too.

The left panels of Fig.~\ref{Fig:HeavyCl} report the average mass number and the $\langle Z \rangle/\langle A \rangle$ value of ``heavy" nuclei produced in neutron rich matter ($Y_p=0.1$) with $0.1 \leq T \leq 5.75~\mathrm{MeV}$ along with the mass fraction which this category binds. 
Shell effects are visible up to $T=0.58~\mathrm{MeV}$. As already discussed by \cite{Hempel_NPA_2010,Gulminelli_PRC_2015,Raduta_NPA_2019} among others, ``heavy" clusters' size decreases with temperature and increases with density; deviations from the latter trend appear only at low temperatures and high densities because of structure effects.
$X_{\mathrm{heavy}}$ features the same behavior as $\langle A_{\mathrm{heavy}}\rangle$. At high densities and low (high) temperatures, its values are about 40\% (10\%).
As the temperature increases, also $\langle Z_{\mathrm{heavy}} \rangle/\langle A_{\mathrm{heavy}} \rangle$ increases; this is due to increasing abundance of ligher and more isospin symmetric species.
The left panel of Fig.~\ref{Fig:LightCl} illustrates the average mass and charge numbers of the nuclei with $Z \geq 3$ and $A <20$ as well as their mass fraction.
It turns out that these species only appear for $T \gtrsim 0.5~\mathrm{MeV}$. 
As the temperature increases, the density domain where they are present shrinks and moves toward higher densities, while $X_{\mathrm{light}}$ increases. 
Nonetheless, for all thermodynamic conditions considered here, $X_{\mathrm{light}}$ never exceeds a few percents.
The suppression of light clusters at $n_B=10^{-6}~\mathrm{fm}^{-3}$ and $T=0.58~\mathrm{MeV}$ is due to the onset of extra H and He isotopes.
Abundant production at high densities of an increasing number of H and He isotopes, which compete with light clusters, also explains the double-peaked distribution of $X_{\mathrm{light}}$ for $T=3.63$ and 5.75 MeV.

The right panels of Fig.~\ref{Fig:HeavyCl} demonstrate the same quantities plotted in the left panels but for $T=1~\mathrm{MeV}$ and $0.01 \leq Y_p \leq 0.6$. For all $Y_p$ values, nuclear shell and odd-even effects make that $\langle A_{\mathrm{heavy}} \rangle$ as a function of $n_B$ presents a sequence of plateaus.
The same is true for $\langle Z_{\mathrm{heavy}} \rangle$ (not shown).
This structure appears most clearly for $Y_p=0.5$. For $Y_p=0.4$, $\langle A_{\mathrm{heavy}} \rangle$, $\langle Z_{\mathrm{heavy}} \rangle$
and $X_{\mathrm{heavy}}$ reach the largest values; in particular, over $10^{-4}~\mathrm{fm}^{-3}\lesssim n_B \leq n_{\mathrm{tr}} \approx 6.7 \cdot 10^{-2}~\mathrm{fm}^{-3}$, $X_{\mathrm{heavy}} \approx 1$.
For $n_B \lesssim 10^{-6}~\mathrm{fm}^{-3}$ and $0.1 \leq Y_p \leq 0.6$, the clusters are almost symmetric; this is due to the fact that they are relatively light. For higher $n_B$, $\langle Z_{\mathrm{heavy}} \rangle/\langle A_{\mathrm{heavy}} \rangle$ decreases with $n_B$ and increases with $Y_p$.
The former feature is due to the fact that $\langle A_{\mathrm{heavy}} \rangle$ increases; the latter is the result of more protons being available.
Remarkably, for high  $n_B$ and $0.4 \leq Y_p \leq 0.6$, $\langle Z_{\mathrm{heavy}} \rangle/\langle A_{\mathrm{heavy}} \rangle \approx Y_p$.
The right panels of Fig.~\ref{Fig:LightCl} show that the largest values of $\langle A_{\mathrm{light}} \rangle \approx 17$
are obtained for $0.01 \leq Y_p \leq 0.25$ and $n_l \lesssim n_B \lesssim n_{\mathrm{tr}}$; for $Y_p=0.01$ (0.25), $n_l=2.5 \cdot 10^{-3}~\mathrm{fm}^{-3} $  ($1.6 \cdot 10^{-2} ~\mathrm{fm}^{-3}$).
The largest values of $\langle Z_{\mathrm{light}} \rangle \approx 6$ are obtained around $n_B=10^{-5}~\mathrm{fm}^{-3}$ for $Y_p=0.5,~0.6$.
Structure effects and/or competition among species lead to irregularities in the $n_B$-dependence of these quantities.
Similarly to the situations considered in the left panel, $X_{\mathrm{light}} \lesssim 2\%$; the highest values correspond to $Y_p=0.6$.

The relative abundance of unbound nucleons and their proton fraction are investigated in Fig.~\ref{Fig:Gas}.
The left bottom panel shows that both $n_{\mathrm{gas}}/n_B$ and the density domain where the gas of unbound nucleons dominates increase with temperature.
The onset of nuclei causes the gas to become depleted in protons. 
The transition to homogeneous matter manifests as a steep increase in both $n_{\mathrm{gas}}(n_B)/n_B$ and $Y_p(n_B)$; for the considered value of $Y_p$, $n_{\mathrm{tr}}$ is independent of $T$. 
The right bottom panel shows that for $T=1~\mathrm{MeV}$ the onset density of the nuclear clusters does not depend on $Y_p$;
the same holds for higher values of $T$.
In contrast, the density for the transition to homogeneous matter depends on $Y_p$ (even if this is barely visible on the plot).
Almost pure neutron and proton gases exist over $10^{-5}~\mathrm{fm}^{-3} \leq n_B \leq n_{\mathrm{tr}}$ for $Y_p \leq 0.4$ and $Y_p \geq 0.5$, respectively, but
for $0.4 \leq Y_p \leq 0.5$, the unbound nucleon gas represents only a tiny fraction of matter. 

The case of hydrogen and helium isotopes, not included in the ``light" nuclei, is analyzed in Fig.~\ref{Fig:HandHe} for $T=1.44~\mathrm{MeV}$.
The top panel illustrates the $n_B-Y_p$ domains where the mass fraction of different hydrogen isotopes exceeds 0.2\% as well as the values of these mass fractions.
It comes out that only three species are populated abundantly enough: $^2$H, $^3$H and the very exotic $^7$H.
Significant production of the isospin symmetric $^2$H occurs in matter with $n_B \approx 2 \cdot 10^{-6}~\mathrm{fm}^{-3}$ and $Y_p \gtrsim 0.25$;
the more neutron-rich $^3$H requires densities higher by a factor of ten and $Y_p \lesssim 0.35$;
neutron-rich dense matter favors the production of extremely neutron-rich $^7$H, which binds up to 20 times more mass than each of the lighter isotopes.
The bottom panel illustrates the $n_B-Y_p$ domains where the mass fractions of neutron-rich isotopes of helium exceed 1\%.
It turns out that dense neutron-rich matter also favors the production of $^{11}$He, $^{12}$He, $^{13}$He, $^{14}$He.
$^{14}$He ($^{11}$He) is present over the widest (narrowest) $n_B-Y_p$ domain.
The highest values of $X_{\mathrm{He}}$ obtained under these circumstances are the following: 1.5\% ($^{11}$He), 3.3\% ($^{12}$He),
9.4\% ($^{13}$He) and 71.2\% ($^{14}$He).

Confirmation of abundant production of light exotic nuclei in dense matter requires a more elaborated treatment of in-medium effects (i.e., beyond excluded volume). So far only the case of $^2$H, $^3$H, $^3$He and $^4$He in symmetric matter has been microscopically addressed~\citep{Typel_PRC_2010}, which is not sufficient for a comprehensive understanding of stellar matter. However, \cite{Hempel_PRC_2011} suggest that, whenever light clusters are abundant, accurate treatment of their energy shifts results in yields larger than those provided by the excluded volume approximation implemented in eNSE.


\subsubsection{Energetics}
\label{sssec:Energetics}

\begin{figure*}
  \centering
  \includegraphics[scale=0.30]{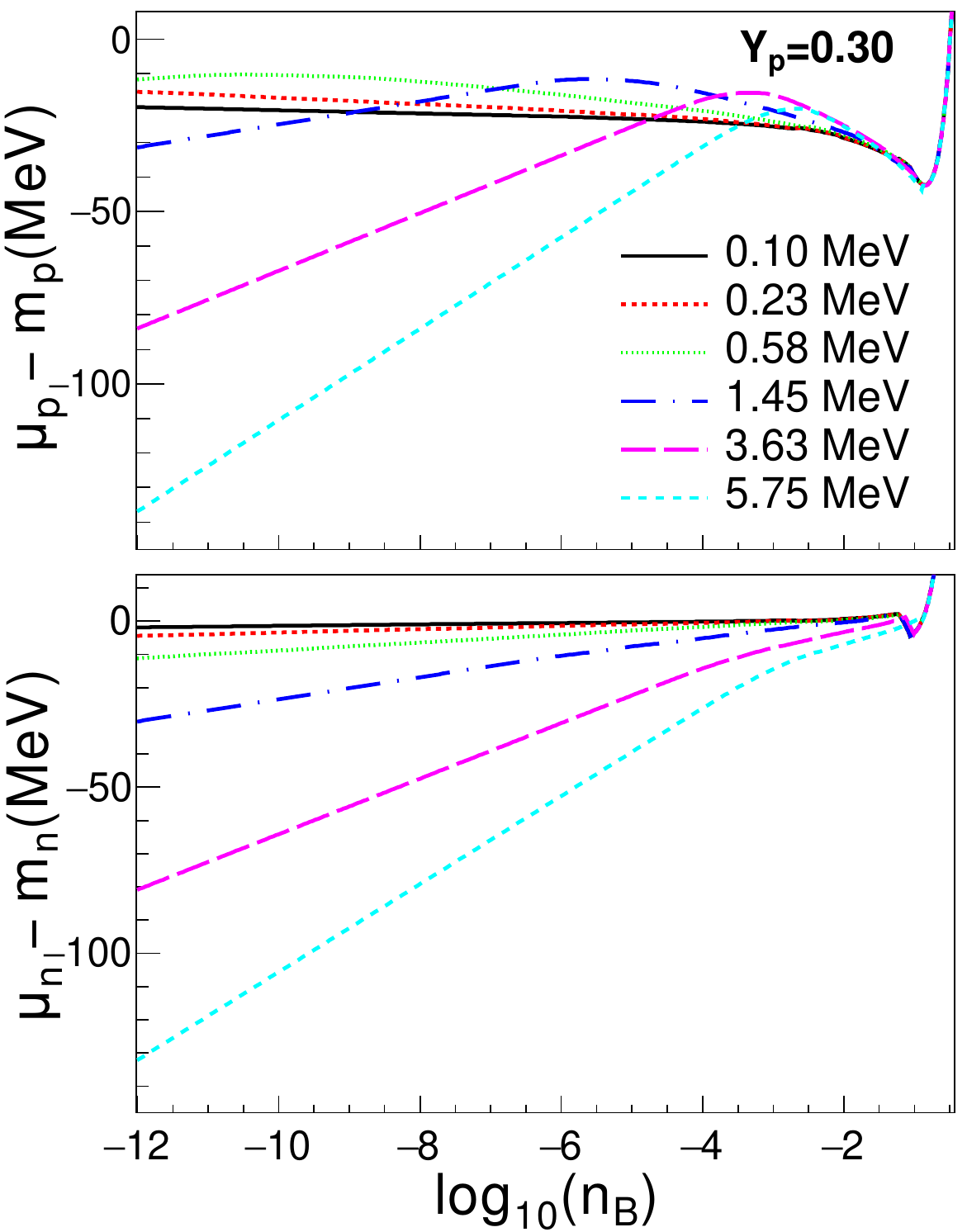}
  \includegraphics[scale=0.30]{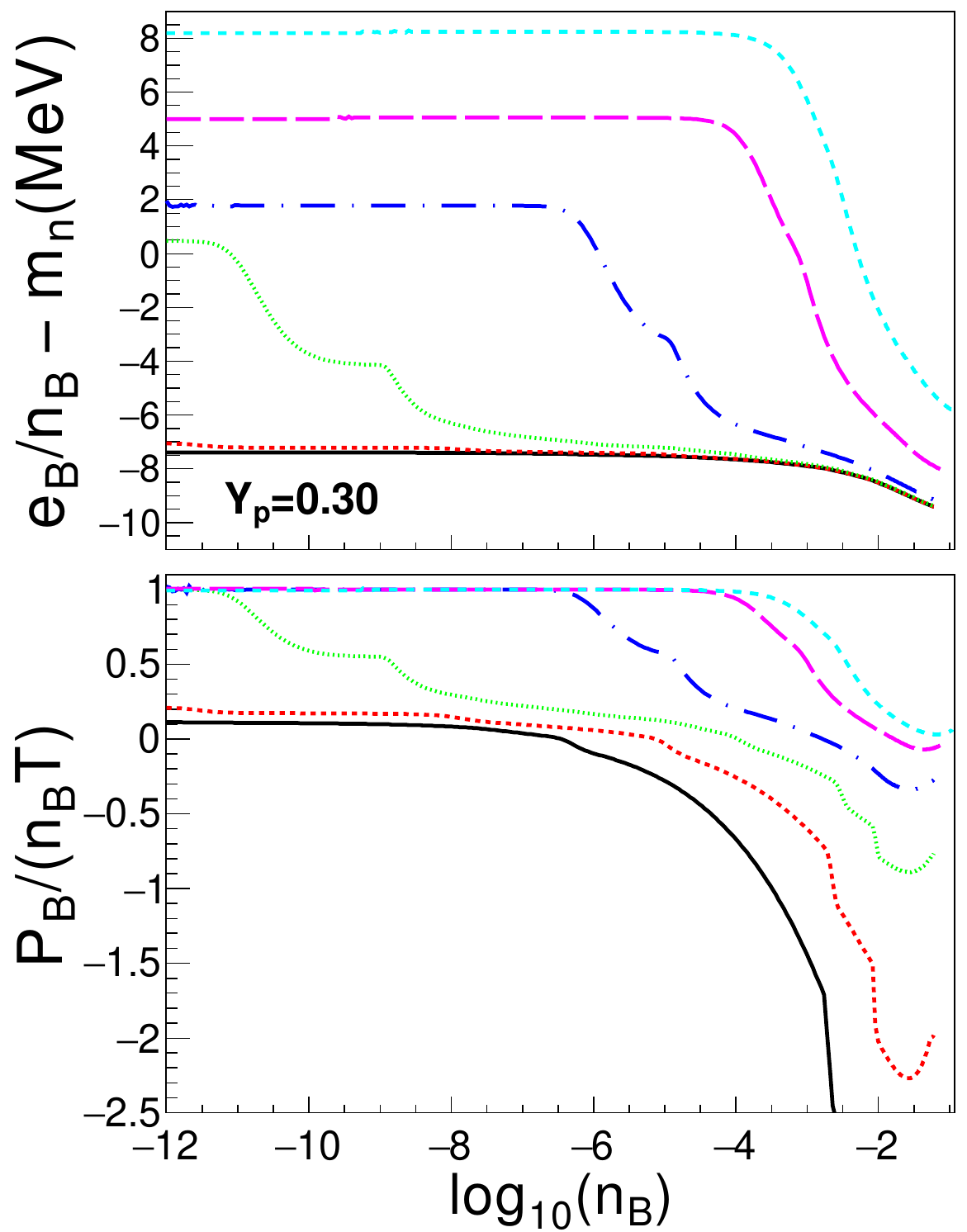}
  \caption{Neutron and proton chemical potentials (left panels), $P_B/(n_B T)$ and energy per particle ($E/A=e_B/n_B$) (right panels)
    as functions of density for $Y_p=0.3$ and $0.1~\mathrm{MeV} \leq T \leq 5.75~\mathrm{MeV}$.
  For better readability, the contributions of electron and photon gases are removed from pressure and energy per particle.
 }
  \label{Fig:Thermo_Yp=0.3}
\end{figure*}

\begin{figure*}
  \centering
  \includegraphics[scale=0.30]{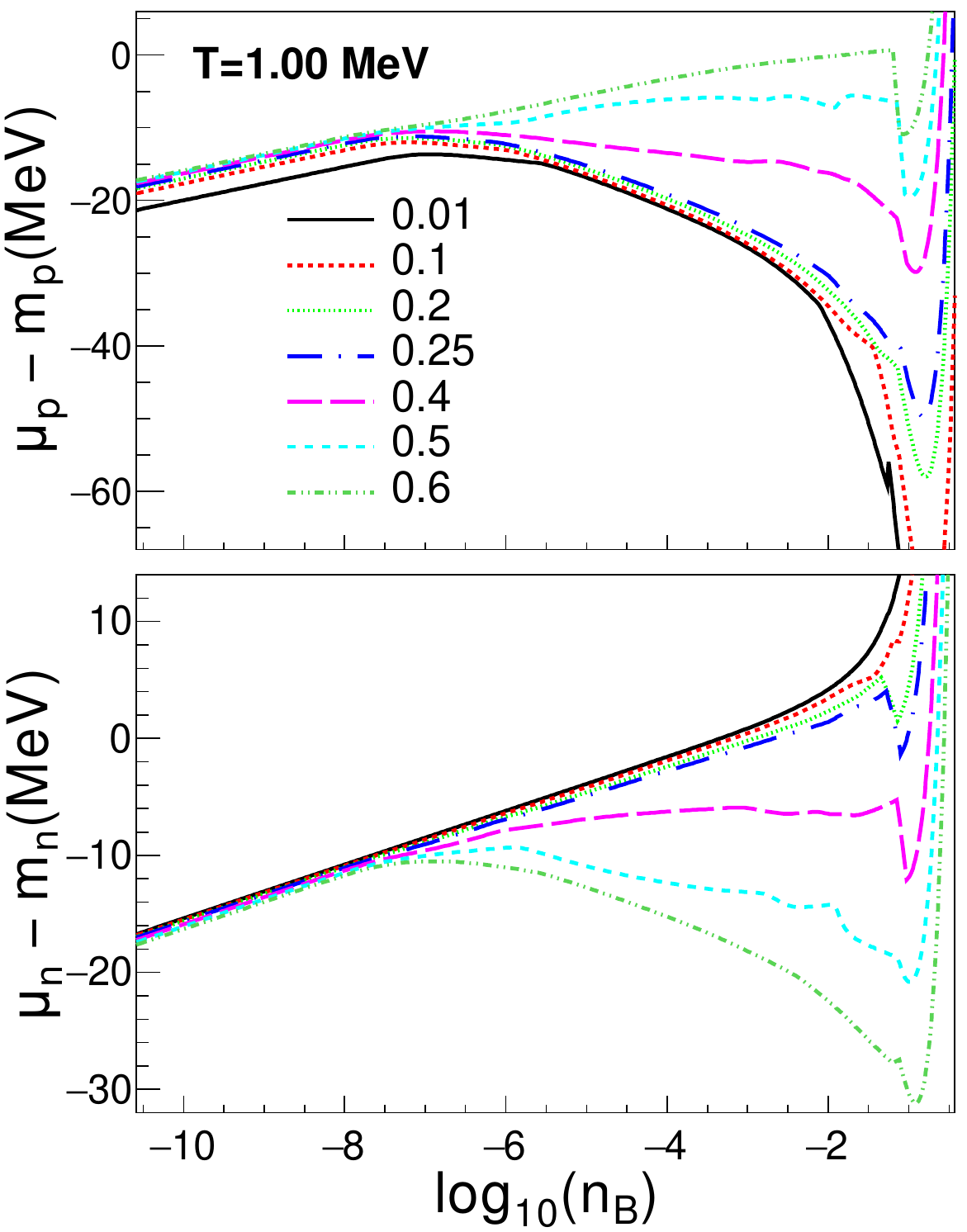}
  \includegraphics[scale=0.30]{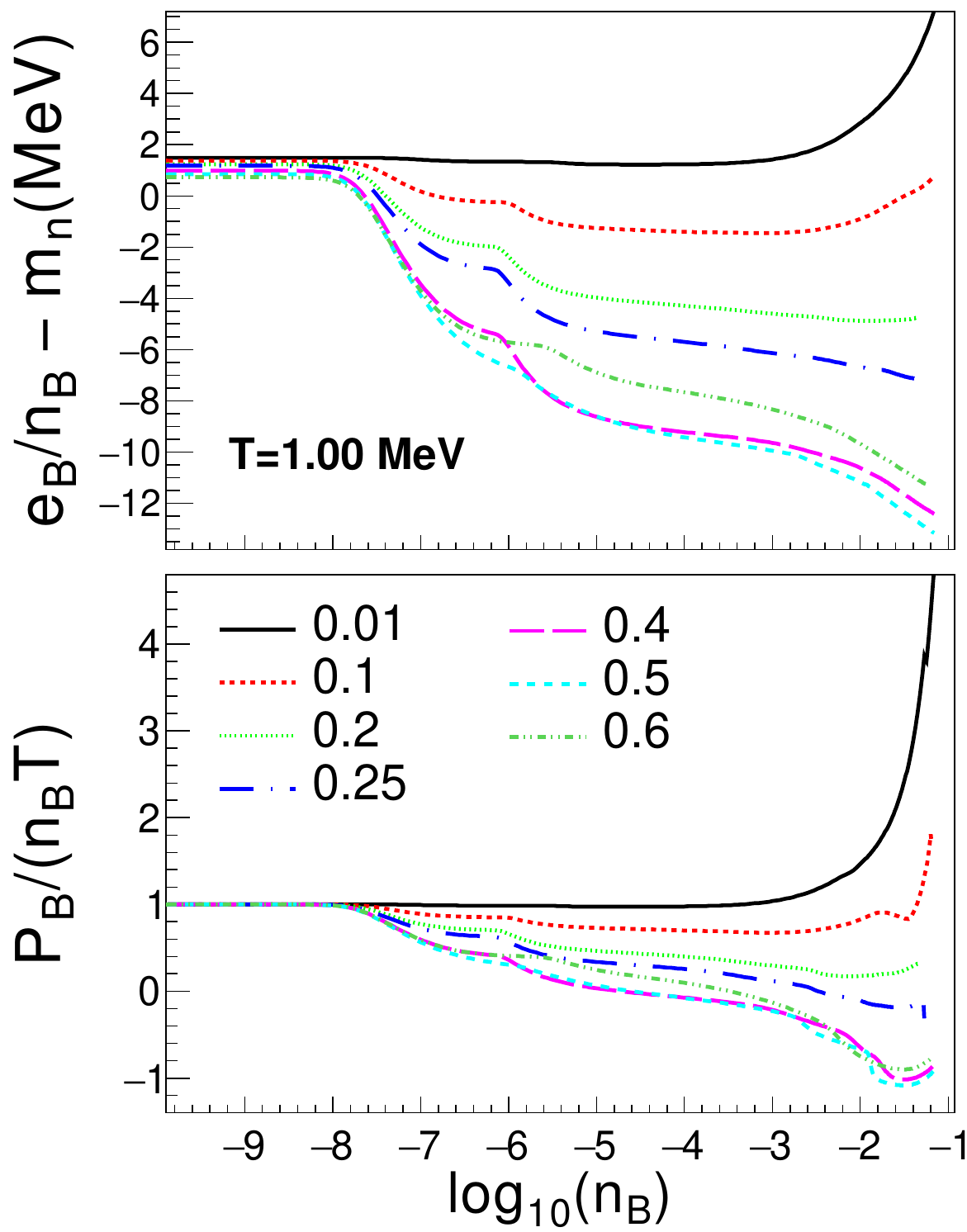}
  \caption{The same as in Fig.~\ref{Fig:Thermo_Yp=0.3} but for constant temperature and various values of $Y_p$.
  }
  \label{Fig:Thermo_T=1}
\end{figure*}

The $T$-, $Y_p$- and $n_B$-dependencies of the energetic state variables are addressed here as functions of density for constant values of $T$ ($Y_p$) and different values of $Y_p$ ($T$).
Only the neutron and proton chemical potentials, (scaled) baryonic pressure, and baryonic energy per particle are considered.
The latter two quantities are obtained upon subtraction of electron and photon gas contributions from the total energy density and total pressure and are preferred for better readability. 
Note that the Coulomb interaction among nuclei and electrons, as well as the one among electrons, is not affected by this subtraction.

Figure~\ref{Fig:Thermo_Yp=0.3} deals with the $T$- and $n_B$-dependencies of matter with $Y_p=0.3$.
For $T \leq 0.58~\mathrm{MeV}$ and $n_B \lesssim 10^{-2}~\mathrm{fm}^{-3}$, both chemical potentials have low values and show little $n_B$-dependence.
This behavior suggests that the unbound nucleon gas is dilute and that much of the density variation of $\mu_{\mathrm{MF};n/p}^0$ is annihilated by the cluster-induced offset $T v_0\sum_{A,Z} N_{A,Z}/V_{\mathrm{cl}}^{\mathrm{free}}$, see Eq.~\eqref{eq:mugas}.
For the two lowest values of $T$, the first conclusion is corroborated by the values of the baryonic energy per particle, $e_B/n_B$, which is consistent with a large fraction of matter bound in clusters, as well as its flatness as a function of $n_B$.
In fact, the value of $e_B/n_B -m_n\approx -7~\mathrm{MeV}$ is due to the binding energy of nuclei, the other terms in Eq.~\eqref{eq:Etot} being negligible. 
For $T=0.58~\mathrm{MeV}$, $e_B(n_B)/n_B$ features a more complex structure that signals a transition from a gas-dominated composition at low densities, with $e_B/n_B-m_n \approx 0.5 ~\mathrm{MeV}$, to a cluster-dominated composition at $n_B=10^{-6}~\mathrm{fm}^{-3}$, where the values of $e_B/n_B$ are very similar to those obtained at lower temperatures.
For $T \geq 1.45~\mathrm{MeV}$ and $n_B \lesssim 10^{-5}~\mathrm{fm}^{-3}$, $\mu_n$ and $\mu_p$ manifest strong $T$- and $n_B$-dependencies; this indicates that the gas component outsizes the clusterized one. Indeed, $e_B/n_B-m_n$ is positive. 
For $10^{-3}~\mathrm{fm}^{-3} \lesssim n_B \lesssim 10^{-1}~\mathrm{fm}^{-3}$, $\mu_p(n_B)$ decreases steeply at all temperatures.
The same is the case of $e_B(n_B)/n_B$. This is the domain preceding the transition to homogeneous matter, where most of the matter is bound in clusters.
Indeed, for $T \leq 3.63~\mathrm{MeV}$,  $e_B/n_B -m_n<0$.
In the proximity of the transition to homogeneous matter, $\mu_n(n_B)$ presents a sudden drop of $\approx 5~\mathrm{MeV}$.
Contrary to what happens in the case of $\mu_p(n_B)$, this is the artifact of the matching between the inhomogeneous and homogeneous phases, see Sec.~\ref{sec:Transition}.
Higher densities correspond to the homogeneous matter regime; there, $\mu_{n,p}(n_B)$ increase steeply and show limited sensitivity to temperature.

With the exception of $T=0.58~\mathrm{MeV}$, $e_B(n_B)/n_B$ and $P_B(n_B)/(n_B T)$ feature wide plateaus over $10^{-12} ~\mathrm{fm}^{-3}\lesssim n_B \lesssim 10^{-6}-10^{-3}~\mathrm{fm}^{-3}$. In the case of low-$T$, this result corresponds to clusterized matter, but the plot does not allow one to understand whether the ideal gas regime is reached or not.
For $T \geq 1.45~\mathrm{MeV}$ and low densities, $P_B/(n_B T)=1$, which corresponds to the ideal gas of unbound nucleons. 
For higher densities, the behavior of $P_B(n_B)/(n_B T)$ versus $n_B$ is very similar to that of $e_B(n_B)/n_B$ versus $n_B$.
We notice that, due to Coulomb, for low $T$ and high $n_B$, $P_B<0$, in agreement with \cite{Hempel_NPA_2010,Raduta_NPA_2019}.
Note that for both $e_B(n_B)/n_B$ and $P_B(n_B)/(n_B T)$ only the behavior of clusterized matter before the matching with homogeneous matter is depicted.
The reason for dropping the higher density region resides in the difficulties of subtracting the electron contributions from the total quantities stored in the tables.

Complementary information is provided in Fig.~\ref{Fig:Thermo_T=1}, where we illustrate the $Y_p$-dependence of $\mu_{n,p}$, $e_B/n_B$ and $P_B/(n_B T)$ as functions of $n_B$ at $T=1~\mathrm{MeV}$.
For $Y_p=0.01$, where less than 0.5\% of the matter is bound in clusters, see Fig.~\ref{Fig:HeavyCl}, the plateaus of $e_B/n_B$ and $P_B/(n_B T)$ as functions of $n_B$ extend over the widest density domain, and both $\mu_{n}(n_B)$ and $\mu_{p}(n_B)$ are smooth.
It turns out that the extremely neutron-rich NM behaves like an ideal gas up to $n_B \approx 10^{-3}~\mathrm{fm}^{-3}$.
For $0.4 \leq Y_p \leq 0.6$, where ``heavy" clusters bind the largest fraction of matter, see Fig.~\ref{Fig:HeavyCl}, $e_B/n_B$ and $p_B/(n_B T)$ reach the lowest negative values. For $Y_p \geq 0.1$, deviations from the ideal gas behavior start to manifest at $n_B=10^{-8} ~\mathrm{fm}^{-3}$.
As before, the sharp drops of $\mu_{n/p}$ around $n_B \approx 6 \cdot 10^{-2} ~\mathrm{fm}^{-3}$ are artifacts of our matching procedure. 

The EOS tables proposed in this paper use the same pool of nuclei and the same definition of clusters. As such, the model dependence between them is only due to different effective interactions. This dependence increases with the isospin asymmetry and barely manifests before the transition to homogeneous matter. It affects mostly the transition density and the composition of matter.
In general, the main source of model dependence comes from the pool of nuclei and the definition of clusters in terms of binding energies, level density, maximum excitation energy, interactions with the medium, evolution of the shell effects with departure from the stability valley, and temperature.
These aspects have been thoroughly addressed in the literature~\citep{Hempel_PRC_2011,Buyukcizmeci_NPA_2013,Furusawa_ApJ_2013,Furusawa_NPA_2017,Furusawa_JPG_2017,Raduta_EPJA_2021}
and are beyond the scope of the present work. 
The behavior of supra-saturated matter, which is of outmost importance for the evolution of astrophysical phenomena, will be discussed elsewhere~\citep{Raduta_inprep_2025}.

\section{Outlook}
\label{sec:Conclusions}

In this paper, we propose a new set of general-purpose EOS tables for simulations of CCSN and BNS mergers.
They are available on \textsc{CompOSE}.
They are derived within the non-relativistic mean-field model of NM and employ Brussels extended Skyrme interactions previously generated within a Bayesian inference of the EOS of dense matter~\citep{Beznogov_PRC_2024}.
The underlying effective interactions were selected in such a way as to manifest the upmost variability in the behavior of Landau effective masses in the suprasaturation domain.

An eNSE model was developed and used to describe the composition and energetics of inhomogeneous matter present at sub-saturation densities and for temperatures up to a dozen MeV. 
The pool of nuclei was represented by the nuclides present in the AME2020~\citep{AME2020} and DZ10~\citep{Duflo_PRC_1995} tables, from which the binding energies were taken. 
Internal partition functions were computed according to the back-shifted Fermi gas level density~\citep{Bucurescu_PRC_2005,vonEgidy_PRC_2005} assuming that the upper limit of the excitation energy is equal to the binding energy.
Interactions among nuclei and with the unbound nucleons were treated within the geometrical excluded-volume approximation. 
The isospin dependence of the intrinsic volumes of the clusters was implemented via the isospin dependence of the saturation density of the NM.

The highly efficient computer code developed to solve the eNSE equations allowed us to study how the transition to homogeneous matter occurs. 
As such, three mechanisms were identified. 
Two of them are possible due to the excluded volume approximation and correspond to the situations in which one of the two components, i.e., the ensemble of nuclei or unbound nucleons, fills the whole available volume. 
The third mechanism corresponds to the case where the homogeneous matter becomes energetically more favorable.
The transition density values range from $\approx \sat{n}/3$ to almost $\sat{n}$, depending on temperature and $Y_p$.
The highest values are obtained for temperatures close to the limiting temperature of the Coulomb instability and $0.4 \lesssim Y_p \lesssim 0.6$, which roughly correspond to the situations where the transition occurs because the clusters fill the entire volume.

The ability of our eNSE code to approach $\sat{n}$ made it possible to revisit the issue of thermodynamic instabilities related to the transition from inhomogeneous to homogeneous matter. 
Given that the domain of phase instability with respect to density fluctuations in NM shrinks with temperature, only the case of cold matter ($T=0.1$~MeV) was considered. 
Our results indicate that no phase instability exists in clusterized matter with electrons. 

The composition and energetics of inhomogeneous NM were systematically investigated over wide domains of temperature, density, and proton fraction. 
As expected, the same general features previously commented on by \cite{Hempel_NPA_2010,Blinnikov_AA_2011,Furusawa_ApJ_2011,Furusawa_ApJ_2013,Furusawa_JPG_2017,Furusawa_NPA_2017,Gulminelli_PRC_2015,Raduta_NPA_2019} were obtained. 
An original result here is the abundant production of neutron-rich isotopes of H and He in neutron-rich environments at finite temperatures, which is due to the inclusion of these species into the NSE pool of nuclei. 
Relying on the fact that in the limit of vanishing temperatures the configurations, which minimize the densities of free energy and energy, coincide, we also considered the composition of $\beta$-equilibrated matter at $T=0.1~\mathrm{MeV}$ as a proxy for cold NS matter, even if this temperature is beyond the validity limit of NSE.
We find that a relatively thick layer made of light neutron-rich nuclides, e.g., $^{14}$He or $^7$H, exists at the bottom of the inner crust. These layers are expected to challenge the occurrence of nuclear pasta phases, affect the transport properties of the inner crusts, and the process of crystallization of NS crusts. However, it remains to be confirmed by dedicated microscopic studies.

The obvious drawback of our eNSE model is linked to the treatment of nuclear clusters at sub-saturation densities.
A more realistic implementation would require consideration of species with much larger mass numbers, typically of the order of a thousand, as well as temperature and in-medium modifications of all species' binding energies.
Temperature effects on clusters' energy density functional will modify both their binding energy and volume; they will also suppress shell effects. 
In-medium modifications of the binding energies are essential for the dissolution of nuclei in dense environments and guarantee a natural transition to homogeneous matter. 
A practical way to account for these effects is via microscopic-inspired parametrizations, as done by \cite{Furusawa_ApJ_2013,Furusawa_JPG_2017,Furusawa_NPA_2017,Typel_JPG_2018}.

\begin{acknowledgments}	
We acknowledge support from the Ministry of Research, Innovation and Digitization, CNCS/CCCDI–UEFISCDI, Project No. PN-IV-P1-PCE-2023-0324 and partial support from Project No. PN 23 21 01 02.
The two authors contributed equally to this work.
\end{acknowledgments}	

\bibliographystyle{aasjournal-hyperref}
\bibliography{NSE.bib}
\end{document}